\documentclass[preprint,showpacs,superscriptaddress,floatfix,pre]{revtex4}

\usepackage{amsmath}
\usepackage[dvips]{graphicx}
\usepackage{epsfig}
\usepackage{tabularx}

\begin{document}

\title{Sensitivity below the standard quantum limit in gravitational wave detectors with Michelson-Fabry-Perot readout }

\author{J.~Belfi}

\affiliation{CNISM-Unit\`a di Siena, Dipartimento di Fisica, Universit\`a di Siena, via Roma
  56, 53100 Siena, Italy}
\author{F.~Marin}
\email[Corresponding author: ]{marin@fi.infn.it}
\affiliation{Dipartimento di Fisica, Universit\`a di Firenze, INFN, Sezione di Firenze, and LENS \\ Via Sansone 1, I-50019 Sesto Fiorentino (FI), Italy}

\date{\today}
\begin{abstract}
We calculate the quantum noise limited displacement sensitivity of a Michelson-Fabry-Perot (MFP) with detuned cavities, followed by phase-sensitive homodyne detection. We show that the standard quantum limit can be surpassed even with resonant cavities and without any signal-recycling mirror nor additional cavities. Indeed, thanks to the homodyne detection, the output field quadrature can be chosen in such a way to cancel the effect of input amplitude fluctuations, i.e., eliminating the force noise. With detuned cavities, the modified opto-mechanical susceptivity allows to reach unlimited sensitivity for large enough (yet finite) optical power. Our expressions include mirror losses and cavity delay effect, for a realistic comparison with experiments. Our study is particularly devoted to gravitational wave detectors and we consider both an interferometer with free-falling mirrors, and a MFP as readout for a massive detector. In the latter case, the sensitivity curve of the recently conceived 'DUAL' detector, based on two acoustic modes, is obtained.
\end{abstract}

\pacs{04.80.Nn, 42.50.Lc, 03.65.Ta, 95.55.Ym}

\maketitle

\section{Introduction}

The sensitivity of interferometers used for the measurement of strain or displacement is commonly referred to a so-called standard quantum limit (SQL), calculated considering independent fluctuations of the radiation pressure acting on the sensing mirrors and of the detected light. Both noise terms are derived from the quantum fluctuations of the electromagnetic field. Several studies, starting at least from the beginning of the 80ies, have shown that an apparatus can beat the SQL, and accurate quantum calculations are in general necessary to find the actual sensitivity. The seminal work by Unruh~\cite{Unruh} shows that the SQL can be surpassed if quantum correlation characterizes the measuring electromagnetic field, and Jaekel and Reynaud~\cite{Jaekel} show that in this case an ultimate limit is imposed by the dissipative part of the mechanical susceptivity. Several schemes have been proposed to reach the goal of a sensitivity beyond the SQL, often using additional optical cavities, such as the quantum locking \cite{Courty}, the local meter \cite{Brag97,Kha02,Dan06}, the detuned signal recycling (studied firstly in Ref.~\cite{Meers88} and later analyzed with a deeper attention to quantum noise in Refs.~\cite{Buonanno,Harms}). It should be noticed that a simple detuning from resonance of an optical cavity allows to rotate the field quadratures \cite{Levenson,Lugiato} and create a correlation between amplitude and phase fluctuations, that are commonly related respectively to radiation pressure and detected field fluctuations \cite{Caves}. This effect is exploited in several proposals of schemes for the generation of ponderomotive squeezing \cite{mancini94,Fabre,Corbitt} and for quantum non-demolition measurements \cite{Heidmann97}. A recent work by Arcizet~\textit{et al.}~\cite{Arcizet} clearly explains that a detuned Fabry-Perot cavity can indeed provide a sensitivity well beyond the SQL, with a frequency behavior very similar to that foreseen for interferometers with signal-recycling mirror~\cite{Buonanno,Harms}.

Most of the mentioned studies have been stimulated by the development of large interferometric detectors of gravitational waves (gw). Recently, a new class of gw detectors has been conceived, based on huge masses kept at cryogenic temperature and called DUAL detectors \cite{Cerdonio,Bonaldi,Marin}. Differently from previous massive cryogenic antennas (such as Weber bars), the DUAL system do not exploit particularly a mechanical resonance of a solid body, but it takes advantage of elastic forces to achieve a useful sensitivity in a wide frequency range. At this purpose, it has to give up to the usual resonant mechanical amplifier, and      
it needs a very sensitive readout. One possibility is using a Michelson interferometer with suitable Fabry-Perot cavities in the two arms (Michelson-Fabry-Perot, MFP)\cite{Marin1}. The readout would be in principle similar to the large gw interferometers like VIRGO\cite{VIRGO} and LIGO\cite{LIGO}, but a more complicated mechanical susceptivity and response function to gw must be taken into account. The definition of SQL is less obvious than in usual interferometric detectors, as well as the possibility to surpass it, and a study of such a system fully including quantum noise is still lacking.

In this article we calculate the sensitivity of a MFP interferometer with detuned cavities. The calculation is very similar to the one described in Ref.~\cite{Arcizet} for a simple cavity, and we find indeed spectral curves well beyond the SQL, with shapes comparable to those typical of signal-recycled interferometers. In addition, we include cavity losses for a better comparison with realistic experimental schemes, we introduce as additional degree of freedom the choice of the final detected field quadrature, and we apply the results to both a standard free-falling masses interferometer and to a DUAL detector.  

\section{Theoretical model}

\subsection{Simple cavity}
\label{sec2a}

Before describing our complete model, we analyze the paradigmatic case of a Fabry-Perot cavity with a movable mirror, neglecting mirror losses and cavity field delay (short cavity regime). Such a calculation is reported in details in Ref.~\cite{Arcizet}, and we only add the choice of the detected field quadrature that can be performed by using a local oscillator with tunable phase. Such simplified scheme is useful to understand the physical meaning of the phenomena that will be observed in the complete system.

We use the semi-classical formalism described in Ref.~\cite{Fabre}, valid in the limit of strong fields, where quantum field fluctuations are treated as classic stochastic variables. 

In the limit of high finesse and nearly resonant conditions, the equation for the cavity field $\alpha$ reads
\begin{equation}
\label{eq1}
(-\gamma+i\psi)\alpha+
\sqrt{2\gamma}\,\alpha_{in}=0
\end{equation}
where $\psi=2k L\,\textrm{Mod}[2\pi]$ is the phase detuning from the closest resonance ($k$ is the laser field wavenumber and $L$ is the cavity length), $2\gamma$ is the input mirror intensity transmission, $\alpha_{in}$ is the input field. The electric fields are normalized such that $|\alpha|^2$ is a flux of photons.

The input/output coupler boundary conditions are
\begin{equation}
\label{eq2}
\alpha_{out}=-\alpha_{in}+\sqrt{2\gamma}\,\alpha \, .
\end{equation}

In a linearized analysis, the general electric field can be considered as a sum of a steady state $\bar{\alpha}$ (which is null for vacuum fields) and the fluctuations $\delta\alpha(t)$ around it. The steady state of the intracavity and reflected fields are respectively
\begin{equation}
\label{eq3}
\bar{\alpha}=\frac{\sqrt{2\gamma}}{\gamma-i\bar{\psi}}\,\bar{\alpha}_{in}   
\end{equation}
and 
\begin{equation}
\label{eq4}
\bar{\alpha}_{out}=\frac{\gamma+i\bar{\psi}}{\gamma-i\bar{\psi}}\,
\bar{\alpha}_{in} \, .
\end{equation}
The steady state of the cavity detuning $\bar{\psi}$ is
\begin{equation}
\label{eq5}
\bar{\psi} = \bar{\psi}_{0}+4\,\hbar k^2 \chi |\bar{\alpha}|^2  \, .
\end{equation}
Here $\bar{\psi}_{0}$ is the cold-cavity detuning (for vanishing laser field) and the last term in Eq.~(\ref{eq5}) is the radiation pressure effect, where $\chi$ is the movable mirror susceptivity.

The linearized fluctuations $\delta\psi(t)$ of $\psi$ around its steady state and the field fluctuations can be written in the Fourier space defining
$\delta\alpha(t)=\delta \tilde{\alpha}(\Omega)e^{-i\Omega t}$, $\delta\alpha^*(t)=\delta \tilde{\alpha}^*(\Omega)e^{-i\Omega t}$ and $\delta\psi(t)=\delta\tilde{\psi}(\Omega)e^{-i\Omega t}$.
The equations for such fluctuations read
\begin{equation}
\label{eq6}
\delta\tilde{\psi}(\Omega) = \delta\tilde{\psi}_{0}+ 4\,\hbar k^2 \chi(\bar{\alpha}^*\delta\tilde{\alpha}+\bar{\alpha}\delta\tilde{\alpha}^*) 
\end{equation}
\begin{equation}
\label{eq7}
(\gamma - i\bar{\psi})\delta\tilde{\alpha} - i\bar{\alpha}\delta\tilde{\psi} = 
\sqrt{2\gamma}\,\delta\tilde{\alpha}_{in} 
\end{equation}
where $\delta\tilde{\psi}_0$ is the signal to be detected, and the equation for $\delta\tilde{\alpha}^*$ is the conjugate of Eq.~(\ref{eq7}).

It is useful to use the quadratures of the field fluctuations, defined as
\begin{equation}
\delta p=\delta\tilde{\alpha}+\delta\tilde{\alpha}^* \, \,; \,\, \, \delta q=i(\delta\tilde{\alpha}^*-\delta\tilde{\alpha}) \, .
\end{equation}
$\delta p$ and $\delta q$ correspond respectively to the amplitude and phase fluctuations, referred to the input mean field that is taken as real (i.e., $\bar{\alpha}_{in}^* = \bar{\alpha}_{in}$).

Using such quadratures, the equation (\ref{eq7}) becomes
\begin{equation}
\label{eq8}
\gamma\,\delta p+\bar{\psi}\,\delta q -i(\bar{\alpha}-\bar{\alpha}^*) \delta\tilde{\psi}=\sqrt{2\gamma}\,\delta p_{in}
\end{equation}
\begin{equation}
\label{eq9}
\gamma\,\delta q-\bar{\psi}\,\delta p -(\bar{\alpha}+\bar{\alpha}^*) \delta\tilde{\psi}=\sqrt{2\gamma}\,\delta q_{in}      \, .
\end{equation}

To write clearer expressions, we define $\Psi = \bar{\psi}/\gamma$ (detuning normalized to the half cavity linewidth) and we use a normalized input laser power $p$ (with the dimensions of a force divided by a length, i.e., the inverse of a susceptivity) defined by
\begin{equation}
p=16 \frac{\hbar k^2 \bar{\alpha}_{in}^2}{\gamma^2}=16 \frac{k P_{in}}{\gamma^2 c} 
\end{equation}
where $P_{in} = \hbar kc\bar{\alpha}_{in}^2$ is the real input power ($c$ is the speed of light). 
With these definitions, the expressions (\ref{eq3}) and (\ref{eq4}) for the fields steady states become
\begin{equation}
\label{eq10}
\bar{\alpha}=\sqrt{\frac{2}{\gamma}}\frac{e^{i\eta}}{\sqrt{1+\Psi^2}}\,\bar{\alpha}_{in}
\end{equation}
\begin{equation}
\label{eq11}
\bar{\alpha}_{out}=e^{2i\eta}\,\bar{\alpha}_{in}
\end{equation}
where $\,\,\eta=\arctan\Psi$.

The quadrature fluctuations, according to Eqs.~(\ref{eq8}) and (\ref{eq9}), are rotated by the same angle as the average field. Indeed, we have inside the cavity
\begin{equation}
\label{eq12}
\left( \begin{array}{c}
\delta p  \\
\delta q \end{array}
\right)=\sqrt{\frac{2}{\gamma}}\frac{1}{\sqrt{1+\Psi^2}}
\left( \begin{matrix} \cos\eta & -\sin\eta \cr \sin\eta & \cos\eta \end{matrix} \right)\,
\left( \begin{array}{c}
\delta p_{in}  \\
\delta q_{in} \end{array}
\right)+\left(\frac{2}{\gamma}\right)^{\frac{3}{2}}\frac{\bar{\alpha}_{in}}{1+\Psi^2}
\left( \begin{array}{c}
-\sin2\eta  \\
\cos2\eta \end{array}
\right)\delta\tilde{\psi}
\end{equation}
where we have used Eq.~(\ref{eq10}) for replacing $\bar{\alpha}$. 

For the reflected fields, we find relations similar to Eq.~(\ref{eq12}), where the rotation angle for the field quadratures is again the same as for the steady state field:
\begin{equation}
\label{eq13}
\left( \begin{array}{c}
\delta p_{out}  \\
\delta q_{out} \end{array}
\right)= 
\left( \begin{matrix} \cos2\eta & -\sin2\eta \cr \sin2\eta & \cos2\eta \end{matrix} \right)\,
\left( \begin{array}{c}
\delta p_{in}  \\
\delta q_{in} \end{array}
\right)+\frac{4}{\gamma}\frac{\bar{\alpha}_{in}}{1+\Psi^2}
\left( \begin{array}{c}
-\sin2\eta  \\
\cos2\eta \end{array}
\right)\delta\tilde{\psi}  \,\,  .
\end{equation}

The equation for $\delta\tilde{\psi}$, in the simplified notation, reads
\begin{equation}
\label{eq14}
\delta\tilde{\psi}=\delta\tilde{\psi}_0+4\,\hbar k^2\chi\sqrt{\frac{2}{\gamma}}
\frac{\bar{\alpha}_{in}}{\sqrt{1+\Psi^2}}[\cos\eta\,\delta p+\sin\eta\,\delta q]
\end{equation}
and, replacing Eq.~(\ref{eq12}) for $\delta p$ and $\delta q$, we obtain  
\begin{equation}
\label{eq15}
\delta\tilde{\psi}=\delta\tilde{\psi}_0+4\,\hbar k^2\chi\frac{2}{\gamma}
\frac{\bar{\alpha}_{in}}{1+\Psi^2}\left[\delta p_{in}- \frac{2}{\gamma}
\frac{\Psi}{1+\Psi^2}\,\bar{\alpha}_{in}\,\delta\tilde{\psi}\right]   \,\, .
\end{equation}

The detected quadrature can be chosen at will by tuning the homodyne angle $w$, and the corresponding fluctuations are 
\begin{equation}
\label{eq16}
\delta E_{out}=\delta p_{out} \cos w+\delta q_{out} \sin w  \,\, .
\end{equation}
We remark that, with respect to the input field, an amplitude detection is obtained for $w=0$ while a phase detection corresponds to $\,w=\pi/2$. On the other hand, with respect to the output field (see Eq.~(\ref{eq11})), $w=2\eta\,$ corresponds to a pure amplitude detection and $\,w=2\eta+\pi/2\,$ to a pure phase detection. With $\,w=2\eta$ we have  $\,\,\delta E_{out}=\delta p_{in}\,\,$  while for $\,\,w=2\eta+\pi/2\,\,$  we obtain
\begin{equation}
\label{eq17}
\delta E_{out}=\delta q_{in}+\frac{4}{\gamma}\frac{\bar{\alpha}_{in}}{1+\Psi^2}\delta\tilde{\psi} \, .
\end{equation}
In general, for a given homodyne phase $\,w$, the detected field fluctuations are
\begin{equation}
\label{eq18}
\delta E_{out}=\delta p_{in}\cos(w-2\eta)+\left(\delta q_{in}+\frac{4}{\gamma}\frac{\bar{\alpha}_{in}}{1+\Psi^2}\delta\tilde{\psi}\right)\sin(w-2\eta)  \,\, .
\end{equation}
Replacing the expression (\ref{eq15}) for $\delta\psi$ in Eq.~(\ref{eq18}), we get
\begin{equation}
\label{eq19}
\delta E_{out}=\delta p_{in}\cos(w-2\eta)+\sin(w-2\eta)\left[\delta q_{in}+
\frac{4}{\gamma}\frac{\bar{\alpha}_{in}}{1+\Psi^2}\frac{\delta\tilde{\psi}_0}{1+A}
+\frac{2}{\Psi}\frac{A}{1+A}\delta p_{in}\right]
\end{equation}
with  
\begin{equation}
A=\frac{\Psi}{(1+\Psi^2)^2}\,p\,\chi  \,\,\, .
\end{equation}

The last term in Eq.~(\ref{eq19}) is the contribution of the radiation pressure noise, proportional to the input field amplitude fluctuations $\,\delta p_{in}$.
For $\,w=2\eta+\pi/2$, the amplitude fluctuations only enter into the detection noise through such radiation pressure term. Changing the detection angle from ($w=2\eta+\pi/2$), the contribution of the phase fluctuations $\,\,\delta q_{in}\,$ to the signal-to-noise ratio (SNR) remains the same (both $\,\delta q_{in}\,$ and $\,\delta\psi_0\,$ multiply the angle-dependent factor $\,\sin(w-2\eta)\,\,$). On the other hand, a further contribution of the amplitude fluctuations comes into play, that can compensate the radiation pressure fluctuations. The optimal SNR is obtained when the contribution of $\,\,\delta p_{in}\,$ is completely cancelled, a situation occurring for an angle $\,w_{opt}\,$ such that
\begin{equation}
\tan(w_{opt}-2\eta)=-\frac{1+A}{A}\,\frac{\Psi}{2}  \,\, .
\end{equation}
In this case, the SNR is only limited by the phase fluctuations, and it increases with the coefficient of $\,\psi_0\,$ within square brackets in Eq.~(\ref{eq19}). In particular, the sensitivity increases for negative detuning (i.e., for $A<0$) and one can get in principle unlimited SNR if the denominator of this coefficient vanishes, i.e., for $\,\,A\to-1.\,\,$ This can happen for large enough power and/or susceptivity, more precisely if $\,\,p\chi>16/3\sqrt{3}\,.\,$ Such increase in sensitivity at negative detuning is usually interpreted as due to a modified effective susceptivity (`optical spring`) originated by the position-dependent radiation pressure force.

\subsection{Complete system}

\begin{figure}[t]
\begin{center}
\includegraphics*[width=0.9\columnwidth]{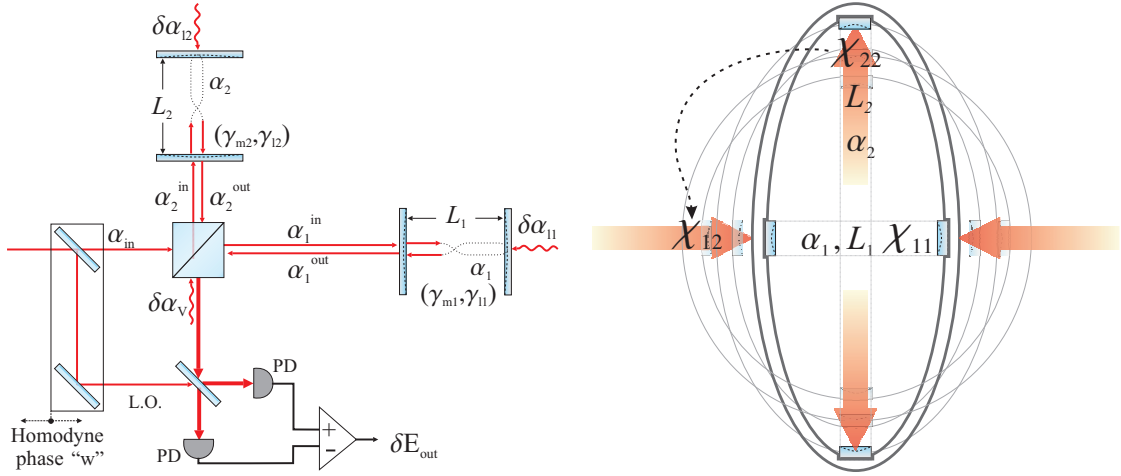}
\end{center}
\caption{Left: the optical configuration considered is a Michelson-Fabry-Perot with the additional free choice of the detected field quadrature. Right: mechanical scheme, with self- and cross- susceptivities.}
\label{schema1}
\end{figure}

The optical scheme we consider, reported in Fig.~(\ref{schema1}), is a MFP with the addition of an homodyne balanced detection at the output. In this way, one can choose the quadrature of the output field to be detected as in the paradigmatic case just analyzed. Vacuum fluctuations are introduced through the mirrors losses in the cavities and through the output port of the Michelson beam splitter, while laser field fluctuations enter through the beam splitter input port. 

Concerning the mechanics of the system, we consider a susceptivity matrix to include the possibility of changing the length of one arm by acting with a force on the other arm. Such a possibility is important in the case of an interferometer mounted on a solid body. A sketch of the mechanical scheme is included in Fig.~(\ref{schema1}) (right).

The equations for the electric fields in the two cavities (labeled by $i,j=(1,2)$) read
\begin{equation}
\label{dadt}
\tau_i \frac{d\alpha_i}{dt}=-(\gamma_i-i\psi_i)\alpha_i+
\sqrt{2\gamma_{mi}}\alpha^{in}_i+\sqrt{2
\gamma_{li}}\delta\alpha_{li}
\end{equation}
where $\tau_i=2L_i/c$ is the cavity roundtrip time, c is the speed of light, $2\gamma_{mi}$ is the input mirror intensity transmission, $2\gamma_{li}$ are the roundtrip intensity losses (including transmission from the back mirror, absorption and scattering in both mirrors, diffraction losses, etc.),  
$\gamma_i=\gamma_{mi}+\gamma_{li}$, $\delta\alpha_{li}$ are the vacuum fluctuations entering through cavity losses which are mimic by a partially transmitting output mirror. 

Assuming an ideal 50$\%$ beam splitter, the input fields $\alpha^{in}_i$ of the two cavities are
\begin{equation}
\label{ain1}
\alpha^{in}_1=\frac{\alpha_{in}-\delta\alpha_V}{\sqrt{2}}
\end{equation}
\begin{equation}
\label{ain2}
\alpha^{in}_2=\frac{\alpha_{in}+\delta\alpha_V}{\sqrt{2}}
\end{equation}
where $\alpha_{in}$ is the laser input field and $\delta\alpha_V$ are the vacuum fluctuations entering through the beam splitter output port. In Eqs.~(\ref{ain1}-\ref{ain2}), to simplify the notation, we have neglected the phase difference between $\alpha^{in}_1$ and $\alpha^{in}_2$ introduced by the length difference in the paths from the beam splitter to the two input cavity mirrors. This phase difference will be re-considered at the output port of the beam splitter (Eq.~(\ref{BSout})).

The cavity input/output coupler boundary conditions are
\begin{equation}
\label{aout}
\alpha^{out}_i=-\alpha^{in}_i+\sqrt{2\gamma_{mi}}\alpha_i \, .
\end{equation}
At the output port of the beam splitter, fields are recombined giving:
\begin{equation}
\label{BSout}
\alpha_{BS}=\frac{1}{\sqrt{2}}(-\alpha^{out}_1+\alpha^{out}_2 e^{i\theta})
\end{equation}
where the phase $\theta$ accounts for the double path difference between the beam splitter and the two input mirrors. Finally, the observed quadrature of the output field can be chosen by changing the detection phase $w$.

The steady state of the intracavity field is obtained by zeroing the time derivative in Eq.~(\ref{dadt}), using Eqs.~(\ref{ain1}-\ref{ain2}) and neglecting the field fluctuations:
\begin{equation}
\bar{\alpha}_i=\frac{\sqrt{\gamma_{mi}}}{\gamma_i-i\bar{\psi}_i}\bar{\alpha}_{in}   \, .
\end{equation}
Using Eq.~(\ref{aout}), we find for the steady state of the reflected field: 
\begin{equation}
\bar{\alpha}^{out}_i=\frac{\gamma_{mi}-\gamma_{li}+i\bar{\psi}_i}{\gamma_{mi}+\gamma_{li}-i\bar{\psi}_i}
\frac{\bar{\alpha}_{in}}{\sqrt{2}} \, .
\end{equation}

The cavity length is sensitive to several kinds of forces acting on the system, including classic deterministic (e.g., the gravitational wave effect), stochastic (thermal noise) and quantum forces (the radiation pressure acting on the mirrors).
The steady state cavity detuning $\bar{\psi}_i$ can be written as 
\begin{equation}
\bar{\psi}_i = \bar{\psi}_{0i}+4\hbar k^2 \chi^0_{ii}|\bar{\alpha}_i|^2+  4\hbar k^2 \chi^0_{ij}|\bar{\alpha}_j|^2
\end{equation}
where $\chi^0_{ij}$ is the stationary (zero-frequency) susceptivity matrix.

The equation for the fluctuations of the cavity phase detuning $\delta\tilde{\psi}_i(\Omega)$ (expressed in the Fourier space) is
\begin{equation}
\label{psi}
\delta\tilde{\psi}_i(\Omega) = \delta\tilde{\psi}_{0i}+ 4\hbar k^2 [\chi_{ii}(\Omega)(\bar{\alpha}^*_i\delta\tilde{\alpha}_i+\bar{\alpha}_i\delta\tilde{\alpha}^*_i)+\chi_{ij}(\Omega)(\bar{\alpha}^*_j\delta\tilde{\alpha}_j+\bar{\alpha}_j\delta\tilde{\alpha}^*_j)]
\end{equation}
where $\delta\tilde{\psi}_0$ contains the effects of thermal and external noise, and of the gw signal and $\chi_{ij}(\Omega)$ is the susceptivity matrix.

The complete set of equations for the fields and cavities displacements are reported in Appendix~A. The complete expressions with different parameters for the two cavities are useful for a future numerical analysis of the effects of the asymmetries and of the allowed tolerances in the parameters. However in this work, for a simpler understanding of the physical phenomena, we will take identical cavities with $\gamma_{m1} = \gamma_{m2} = \gamma_m$, $\gamma_{l1} = \gamma_{l2} = \gamma_l$, $\tau_1 = \tau_2 = \tau$.

As a further restriction to our analysis, we will consider that:

a) the Michelson interferometer working point gives a dark fringe at the beam splitter output port. This corresponds to setting equal distances between the beam splitter and the two cavities, i.e., $e^{i\theta}=1$;

b) the two cavities have the same detuning from the laser frequency: $\bar{\psi}_1 = \bar{\psi}_2 = \bar{\psi}$;

c) the mechanical system is symmetric: $\chi_{11}=\chi_{22}=\chi_s$ and $\chi_{12}=\chi_{21}=\chi_c$.

These three conditions determine the cancellation of the effect of the input laser field fluctuations in the output. This requirement is important for a system working in the acoustic frequency range. Indeed, while it is very difficult to reduce the laser field amplitude fluctuations at the shot noise level~\cite{Conti}, for phase fluctuations the reduction to the quantum limit is even more difficult and far from being demonstrated in strong power laser fields.

The field fluctuations $\delta \tilde{E}_{out}$ (seen by the homodyne detection) are described by a vector of coefficients $V_{out}$, multiplying the input fluctuations $X_{in}$:
\begin{equation}
\label{Eout}
\delta \tilde{E}_{out}(\Omega) = V_{out} \cdot X_{in}(\Omega)
\end{equation}
with $X_{in}$ and $V_{out}$ given respectively in equations (\ref{Xin}) and (\ref{Vout}) of the Appendix, where we have used for convenience the quadratures of the field fluctuations.

As already remarked, the coefficients $V_{out}[\delta p_{in}]$ and $V_{out}[\delta q_{in}]$ that multiply the input laser field fluctuations $\delta p_{in}$ and $\delta q_{in}$ are null in the completely symmetric case that we are considering.

Besides the previous defined normalized detuning $\Psi=\bar{\psi}/\gamma$, we use for a more compact notation $\Gamma_m = \gamma_m/\gamma$ (in the case of loss-less cavities, $\Gamma_m = 1$); $\Omega_{cav}=\gamma/\tau$ (cutoff angular frequency of the cavity) and the normalized input laser power $p$ is now
\begin{equation}
p=16 \frac{\Gamma_m^{3/2}k P_{in}}{\gamma^2 c} \, .
\end{equation}
The expression of the sensitivity $S_L(\Omega)$ in the detection of $\delta(L_1-L_2)$ (defined as the signal spectral power with unitary signal-to-noise spectral density) is given in Appendix~A, Eq.~(\ref{NSsimm}). We see that in $S_L(\Omega)$ the susceptivities only appear as a difference $\chi_s - \chi_c$. We define in the following $\chi_s - \chi_c = \chi$.

\section{General discussion}

The first situation that we consider is with the cavities at resonance ($\bar{\psi} = 0$) and pure phase quadrature detection ($w = \pi/2$). Such configuration corresponds to the present operative gw interferometers VIRGO and LIGO, and we will define it in the following as 'normal case'. The $S_L(\Omega)$ of Eq.~(\ref{NSsimm}) becomes:
\begin{equation}
S_L = 
\frac{\hbar}{\sqrt{\Gamma_m}}\frac{1+(\frac{\Omega}{\Omega_{cav}})^2}{p}\left[1+\left(\frac{p|\chi|}{1+(\frac{\Omega}{\Omega_{cav}})^2}\right)^2\right]  \, .
\end{equation}
It can be seen that $S_L$ can be written as a sum of a 'displacement noise' term $S_{xx}$ and a term proportional to a 'force noise' $S_{FF}$:
\begin{equation}
S_L = S_{xx} + \left|\chi\right|^2 S_{FF}
\end{equation}
with
\begin{equation}
S_{xx} S_{FF} = \hbar^2/\Gamma_m \, .
\end{equation}
The origin of the two terms can be found respectively in the intracavity field phase noise and amplitude noise. The first term limits the detection sensitivity  of a phase signal created by the mirrors displacement; the second term is due to the fluctuations of the radiation pressure acting on the cavity mirrors. For each detection frequency $\Omega$, the optimal $S_L(\Omega)$ is reached for $S_{xx}= \hbar\left|\chi\right|/\sqrt{\Gamma_m}$. With this condition, $S_L$ is equal to what we call standard quantum limit:
\begin{equation}
\label{SQL} 
SQL(\Omega) = \frac{2\hbar\left|\chi(\Omega)\right|}{\sqrt{\Gamma_m}} \, .
\end{equation}
With long cavities (i.e., when $\Omega/\Omega_{cav}$ is not negligible) and/or a frequency-dependent susceptivity, the $SQL$ defines an envelop of possible spectral density curves, each one determined by the choice of the input power.

Already with resonant cavities, tuning the homodyne phase $w$ allows to change the sensitivity significantly. $S_L$ still assumes a simple form:
\begin{eqnarray}
S_L(\Omega)&=& \frac{\hbar}{\sqrt{\Gamma_m}}\frac{1+(\frac{\Omega}{\Omega_{cav}})^2}{p}\Big[1+(
1-\Gamma_m)\left(\frac{p 
Re\chi}{1+(\frac{\Omega}{\Omega_{cav}})^2}\right)^2\\\nonumber&+&\left( 
\frac{\cos w}{\sin{w}}+\sqrt{\Gamma_m}\frac{p 
Re\chi}{1+(\frac{\Omega}{\Omega_{cav}})^2}\right)^2+\left(\frac{p Im\chi}{1+(\frac{\Omega}{\Omega_{cav}})^2}\right)^2\Big]     
\label{NSresonant}
\end{eqnarray}
that, for loss-less cavities ($\Gamma_m = 1$) and real $\chi$, can be further simplified to
\begin{equation}
S_L(\Omega)=
\hbar\frac{1+(\frac{\Omega}{\Omega_{cav}})^2}{p}\left[1+\left(\frac{\cos w}{\sin{w}}+\frac{p \chi}{1+(\frac{\Omega}{\Omega_{cav}})^2}\right)^2\right]    \, .
\label{NSreslossless}
\end{equation} 
The physical interpretation of Eq.~(\ref{NSreslossless}) is simple, as already explained for a simple cavity. Like in the 'normal case', for resonant cavities there is no quadrature rotation, the vacuum fluctuations $\delta p_V$ and $\delta q_V$ are transferred respectively to the amplitude ($\delta p_{BS}$) and phase ($\delta q_{BS}$) fluctuations of the output field, and the radiation pressure is proportional simply to $\delta p_V$. The cavity length fluctuations, besides the signal, contain a term due to the radiation pressure, and a term proportional to the length fluctuations is present in the output field phase. As a consequence, the amplitude fluctuations $\delta p_V$ are found both in the output field amplitude quadrature $\delta p_{BS}$ (giving the first term in round brackets of Eq.~(\ref{NSreslossless})) and in the output phase $\delta q_{BS}$ (second term in round brackets of Eq.~(\ref{NSreslossless})). A suitable choice of the homodyne phase, and therefore of the detected quadrature, brings these terms to cancel each other. Also for $\Gamma_m \ne 1$, choosing
\begin{equation}
w = w_0 = \arctan\left(-\frac{1+(\Omega/\Omega_{cav})^2}{\sqrt{\Gamma_m}\,p\, Re\chi}\right)  \, ,
\label{w0}
\end{equation}
the contribution of amplitude fluctuations is minimized and the $S_L$ becomes
\begin{equation}
S_L = S_L^0 = S_{xx}+S_{FF} [(1-\Gamma_m) Re\chi^2 + Im\chi^2]  \, .
\end{equation}
For loss-less cavities ($\Gamma_m = 1$) and real $\chi$, the radiation pressure noise can be completely cancelled and we have $S_L^0 = S_{xx}$. 

For frequency-dependent susceptivity and/or long cavities, $S_L^0$ defines the locus of the minima of spectral curves that can be tuned by changing $w$. The value of $w$ necessary to achieve the best sensitivity at a chosen frequency is given by Eq.~(\ref{w0}). We remark that in the 'normal case' the $SQL$ is an absolute limit, while in the general resonant case (in particular, for $\Gamma_m = 1$) the sensitivity limit given by $S_{xx}$ can be decreased at will by increasing the laser power.

For detuned cavities, both the radiation pressure and the output field fluctuations contain a mixture of $\delta p_V$ and $\delta q_V$ and the force noise is not due any more solely to the pure amplitude fluctuations $\delta p_V$. A clear discussion of the phenomena involved in this case is given in Ref.~\cite{Arcizet}. 

\begin{figure}[t]
\begin{center}
\includegraphics*[width=0.9\columnwidth]{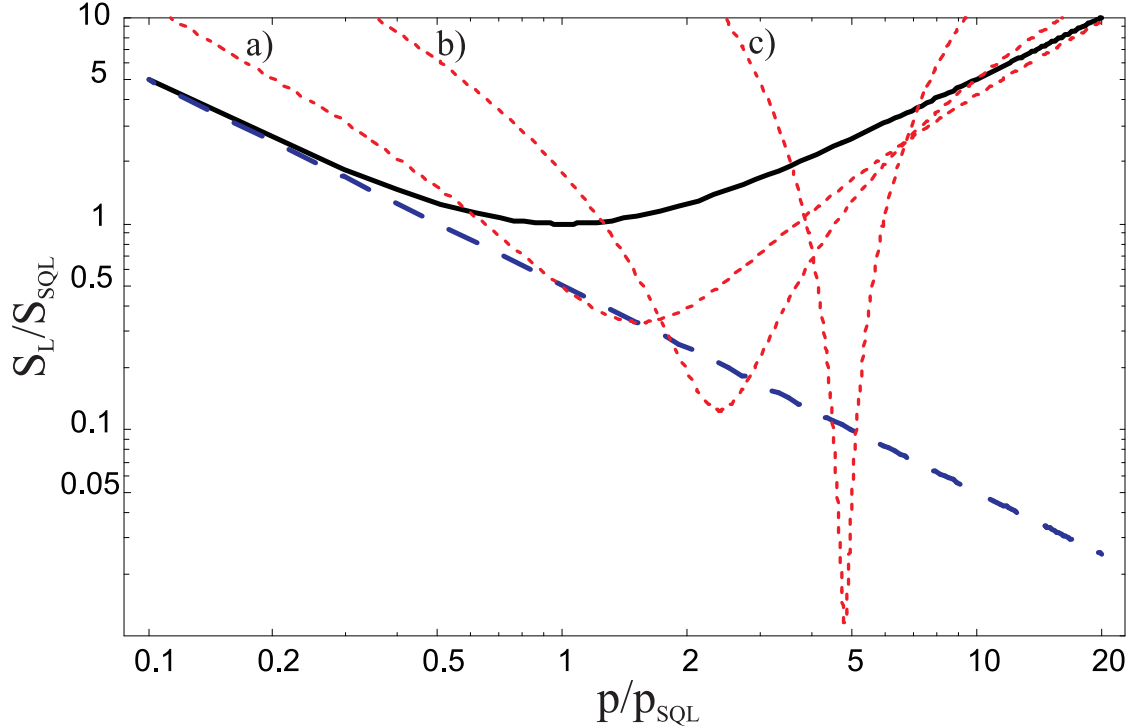}
\end{center}
\caption{Sensitivity $S_L$ as a function of input power $p$, for $\Omega_{cav}\to\infty$, $\Gamma_m=1$, and constant, real $\chi$. Solid line: 'normal case' ($\Psi=0$, $w=\pi/2$); dashed line: general resonant case ($\Psi=0$, $w=\arctan(-1/p\chi)$); dotted lines: $\Psi=-0.4$ and $w=\pi/2$ (a), $w=2.0$ (b), $w=2.3$ (c).}
\label{costantchi}
\end{figure}

For a better understanding of the physics, we analyze the case of a constant, real, positive $\chi$ and very short cavities ($\Omega_{cav} \gg \Omega$) with negligible losses ($\Gamma_m \simeq 1$). The 'normal case' sensitivity is shown in Fig.~(\ref{costantchi}) with a solid line, as a function of the input power. The $SQL$ is reached for $p=p_{SQL}=1/\chi$. For $p > p_{SQL}$ the sensitivity is worse due to strong radiation pressure effect, for $p < p_{SQL}$ it is deteriorated by phase noise. 
The sensitivity $S_L^0$ for the general resonant case is shown in Fig.~(\ref{costantchi}) with a dashed line. In this case, $S_L^0$ coincides with $S_{xx}$. For low power it approaches the 'normal case' sensitivity (that is here dominated by $S_{xx}$), but for $p > p_{SQL}/2$ it surpasses the $SQL$.
If we allow for different values of the detuning $\Psi$, $S_L$ can decrease well below the $SQL$ and even below $S_L^0$, as shown in Fig.~(\ref{costantchi}) with dotted lines, and it is unlimited if $p$ is strong enough. The physics behind this effect was previously explained in Section~\ref{sec2a}. In short terms, it is the result of: a) the cancellation of the input amplitude fluctuations (thanks to a good choice of the homodyne detection angle); b) the modified effective susceptivity, that increases the sensitivity to the mirror motion.  

Considering now a complex susceptivity $\chi$, taking into account mechanical dissipation, it can be shown that the minimal sensitivity is $\,\,\hbar \, Im\chi$, as already found in Ref.~\cite{Jaekel} for a MFP with resonant cavities and squeezed input fields, and by Arcizet~\textit{et al.}~\cite{Arcizet} for a detuned Fabry-Perot cavity. This phenomenon can be understood as follows. In the detected field, the fluctuations $\delta p_V$ and $\delta q_V$ are present for two different reasons: directly in the field reflected by the interferometer (purely optical effect), and because of the length fluctuations induced by radiation pressure (opto-mechanical effect). As we have already seen, the same situation is found in the case of resonant cavities, but only for $\delta p_V$. The two effects give different linear combinations of $\delta p_V$ and $\delta q_V$, with real coefficients depending on $\Psi$ and $w$ (we are considering $\Omega_{cav}\to\infty$). However, the opto-mechanical effect is mediated by the susceptivity $\chi$. If $\chi$ is real, an appropriate choice of $\Psi$, $w$ and $p$ (and therefore of the coefficients multiplying $\delta p_V$ and $\delta q_V$) can bring to a complete cancellation between the purely optical and opto-mechanical effects. As already explained, the same situation, occuring in the case of resonant cavities only for $\delta p_V$, defines $S^0_L$. However, if $\chi$ has an imaginary component, the radiation pressure fluctuations cannot be completely cancelled. In other words, the total detected fluctuations cannot be completely deleted because of the de-phasing between intracavity intensity changes and cavity length changes introduced by the complex $\chi$. In spite of this interesting physical result, a region with significant imaginary part of the susceptivity is of limited practical interest and in the following we will assume a real $\chi$.

\section{Free-falling mirrors}
   
We apply now our results to find the sensitivity of an interferometer with free-falling mirrors, still with loss-less short cavities. Such a scheme is a good approximation for mirrors suspended to a pendulum with low oscillation frequency. The susceptivity can be written as 
\begin{equation}
\label{chifree}
\chi = - \chi_0 \left(\frac{\Omega_0}{\Omega}\right)^2  
\end{equation}
where $\chi_0$ and $\Omega_0$ are constants.
We remark that a finite zero-frequency susceptivity (and therefore a defined pendulum oscillation frequency) is necessary in the calculation of the steady state and stability of the system, but it can be neglected in the evaluation of the spectra. 

\begin{figure}[t]
\begin{center}
\includegraphics*[width=0.9\columnwidth]{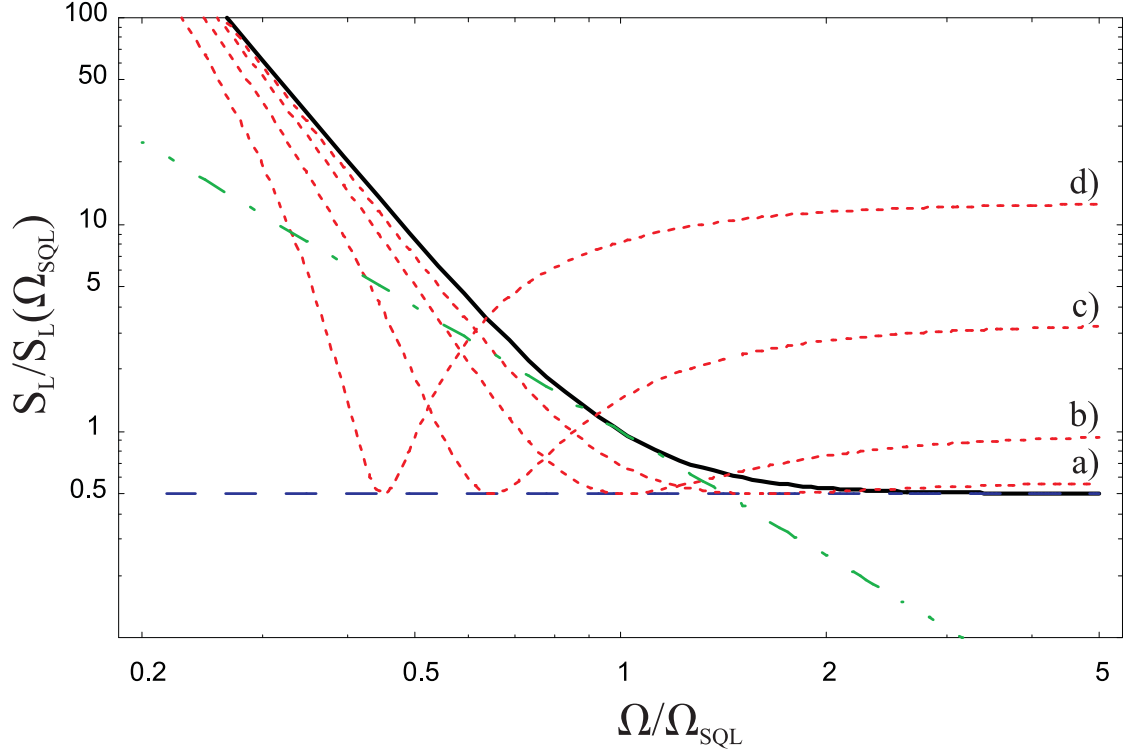}
\end{center}
\caption{Sensitivity $S_L$ in the case of free-falling mirrors, as a function of the frequency $\Omega$, for $\Omega_{cav}\to\infty$, $\Gamma_m=1$, $p=1$. Solid line: 'normal case' ($\Psi=0$, $w=\pi/2$); dashed line: resonant case limit $S_L^0$; dash-dotted line: SQL; dotted lines: $\Psi=0$ and $w=1.2$ (a), $w=0.8$ (b), $w=0.4$ (c), $w=0.2$ (d).}
\label{free1}
\end{figure}

As well known, in the 'normal case' for each particular choice of the power $P_{in}$ the $SQL$ is reached at a corresponding frequency $\Omega_{SQL}$ given by
\begin{equation}
\Omega_{SQL}^2=p\chi_0\Omega_0^2 \, , 
\end{equation}
where $S_L$ is: $\,S_L(\Omega_{SQL})=2\hbar/(\sqrt{\Gamma_m}\,p)$. At high frequencies, $S_L$ tends to the asymptotic value $\,S_L(\Omega\to\infty)=\frac{1}{2}S_L(\Omega_{SQL})$, limited by phase noise, while below $\Omega_{SQL}$ the $S_L$ increases as $\Omega^{-4}$ due to radiation pressure fluctuations. An example is shown in Fig.~\ref{free1} (solid line).

Keeping resonant cavities but changing $w$, the sensitivity curves can surpass the $SQL$, with minima lying on the horizontal line given by $S_{xx}$ (Fig.~\ref{free1}). We remark that in our scheme (with completely symmetric cavities), the average field at the output port vanishes, therefore the local oscillator power can be kept low. At first order, the balanced homodyne detection scheme is not sensitive to the local oscillator noise. Therefore, the phase tuning element (e.g., a phase mirror or an electro-optic modulator) is not critical for the noise budget. As a consequence, this sensitivity tuning technique can be easier to be implemented than other schemes (e.g., tuning the MFP cavities or a signal recycling mirror).

\begin{figure}[t]
\begin{center}
\includegraphics*[width=0.9\columnwidth]{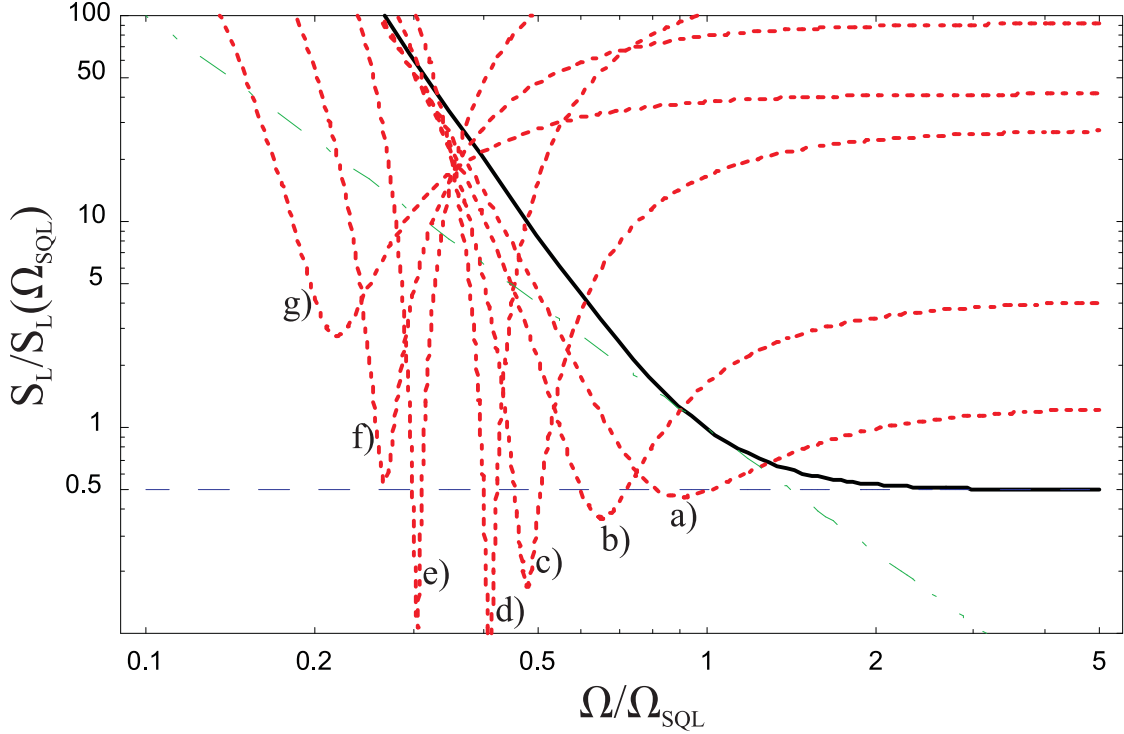}
\end{center}
\caption{Sensitivity $S_L$ in the case of free-falling mirrors, as a function of the frequency $\Omega$, for $\Omega_{cav}\to\infty$, $\Gamma_m=1$, $p=1$. Solid line: 'normal case' ($\Psi=0$, $w=\pi/2$); dashed line: resonant case limit $S_L^0$; dash-dotted line: SQL; dotted lines: $w=\pi/2$ and $\Psi=0.4$ (a), $\Psi=0.6$ (b), $\Psi=0.8$ (c), $\Psi=0.9$ (d), $\Psi=1.1$ (e), $\Psi=1.2$ (f), $\Psi=1.4$ (g).}
\label{free2}
\end{figure}

If we now allow for detuned cavities we see that even $S_L^0$ can be largely surpassed. Some examples are shown in Fig.~(\ref{free2}) for a detection phase $w$ kept at $\pi/2$. We remark that the position of the minimum $S_L$ shifts toward low frequencies at increasing detuning. Therefore, the input power should be increased to keep the same optimal frequency range. Similar results are described in Ref.~\cite{Arcizet} (see their Fig.~3)\cite{nota1}. 

\begin{figure}[t]
\begin{center}
\includegraphics*[width=0.9\columnwidth]{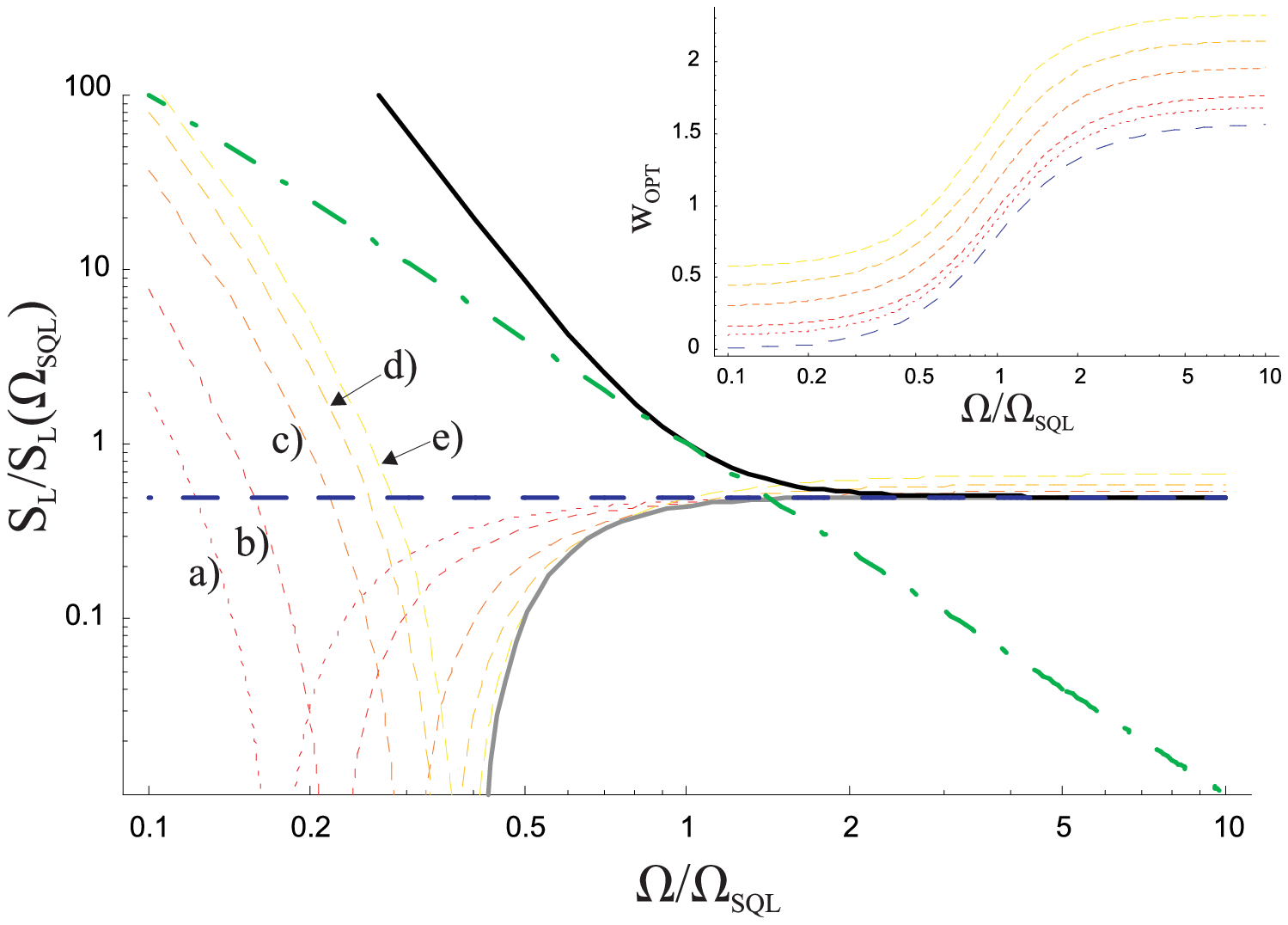}
\end{center}
\caption{Sensitivity $S_L$ in the case of free-falling mirrors, as a function of the frequency $\Omega$, for $\Omega_{cav}\to\infty$, $\Gamma_m=1$, $p=1$. Black solid line: 'normal case' ($\Psi=0$, $w=\pi/2$); long-dashed straight line: resonant case limit $S_L^0$; dash-dotted line: SQL; dashed lines: the phase $w$ is optimized for each frequency, $\Psi=0.06$ (a), $\Psi=0.1$ (b), $\Psi=0.2$ (c), $\Psi=0.3$ (d), $\Psi=0.4$ (e); gray solid line: both $w$ and $\Psi$ are optimized for each frequency. In the inset: phase $w$ optimized for each frequency, for the values of $\Psi$ used in the main plot.}
\label{free3}
\end{figure}

Tuning the phase $w$ allows to obtain families of sensitivity curves, one for each fixed value of detuning, whose envelopes are very broad and low. Some examples are shown in Fig.~(\ref{free3}), where the gray solid curve represents the absolute sensitivity limit that can be reached optimizing, for each frequency, both the phase $w$ and the detuning. In the inset of the figure is shown the optimal phase $w$, as a function of the frequency, that is necessary to obtain the envelope curves. For resonant cavities, inserting the susceptivity (\ref{chifree}) in equation~(\ref{w0}), we find that the optimal phase is  $w=w_0=\arctan(\Omega/\Omega_{SQL})^2$. The other curves differ from $w_0$ for a very flat additional phase. We notice that, besides choosing a particular detection phase to optimize the detection at the preferred frequency, one can also obtain a sensitivity curve corresponding to one of the above envelopes by adding a suitable frequency-dependent quadrature rotation at the output, before the homodyne detection, like in the so-called variational-output interferometers (\cite{Vyatchanin95,Kimble01}).   

\begin{figure}[t]
\begin{center}
\includegraphics*[width=0.9\columnwidth]{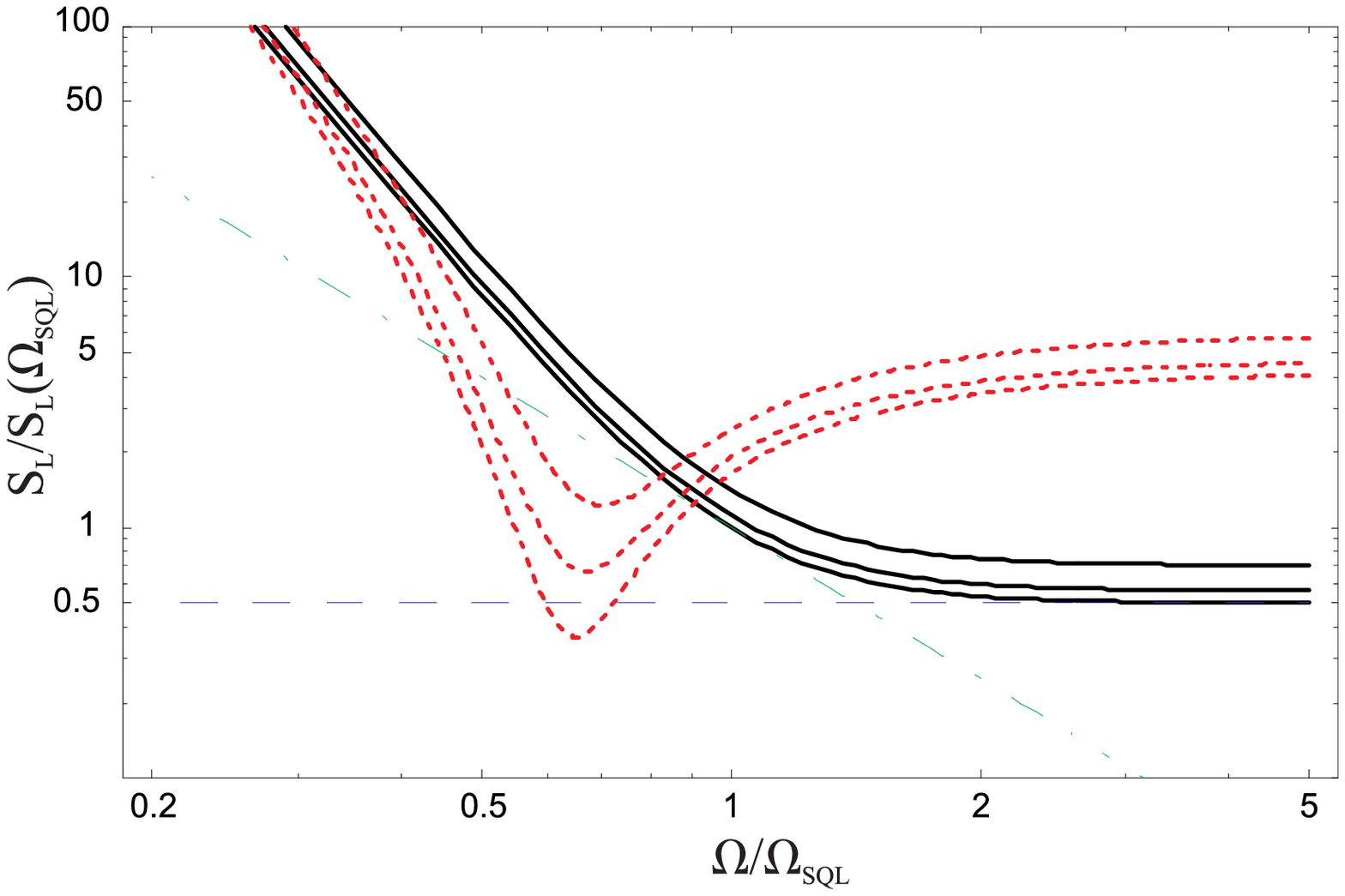}
\end{center}
\caption{Sensitivity $S_L$ in the case of free-falling mirrors, as a function of the frequency $\Omega$, for $\Omega_{cav}\to\infty$, $p=1$. Solid lines: 'normal case' ($\Psi=0$, $w=\pi/2$, from lower to upper curve $\Gamma_m=1,\, 0.8,\, 0.5$); dashed line: resonant case limit $S_L^0$ for $\Gamma_m=0$; dash-dotted line: SQL for $\Gamma_m=0$; dotted lines: $w=\pi/2$, $\Psi=0.6$, and from lower to upper curve $\Gamma_m=1,\, 0.8,\, 0.5$.}
\label{losses}
\end{figure}

Moderate cavity losses do not change qualitatively the above features, and also quantitatively the enhanced sensitivity is well preserved. An example is shown in Fig.~(\ref{losses}). Even at $\Gamma_m = 0.5$ (input mirror transmission equal to the other roundtrip losses) the dip in the $S_L$ is well pronounced, below the $SQL$. 

\begin{figure}[t]
\begin{center}
\includegraphics*[width=0.9\columnwidth]{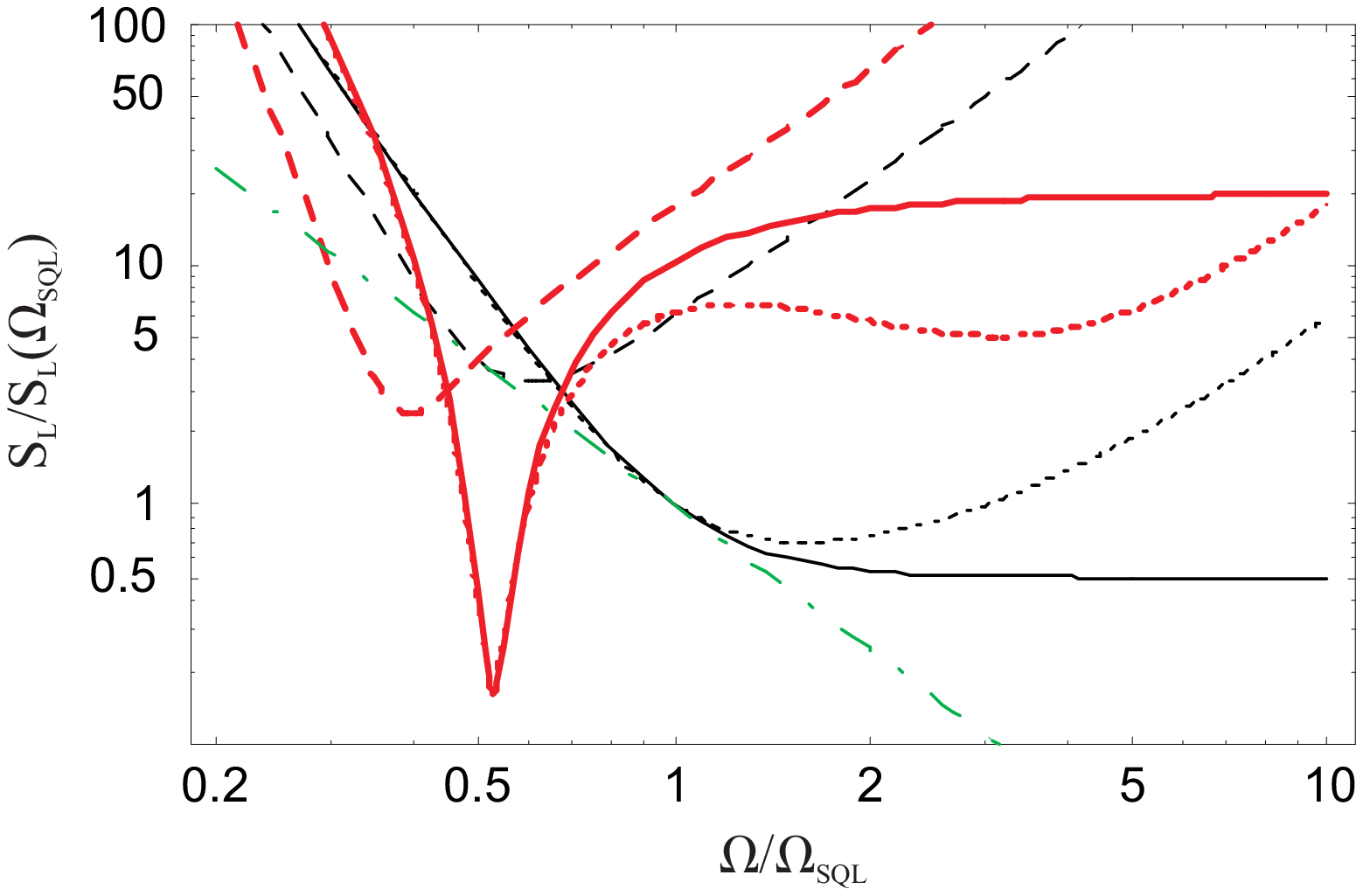}
\end{center}
\caption{Sensitivity $S_L$ in the case of free-falling mirrors, as a function of the frequency $\Omega$, for $\Gamma_m=1$, $p=1$. Thin, dark lines (black online): 'normal case' ($\Psi=0$, $w=\pi/2$); thick, light lines (red online): $\Psi=0.6$, $w=1.3$. Solid lines: $\Omega_{cav}\to\infty$; dotted lines: $\Omega_{cav}=3 \,\Omega_{SQL}$; dashed lines: $\Omega_{cav}=0.3 \,\Omega_{SQL}$. Dash-dotted line: SQL.}\label{cut}
\end{figure}

The expression of $S_L$ including the cavity response is reported in the Appendix~A (Eq.~(\ref{NSsimm})). 
The simplified expression for the 'normal case' is
\begin{equation}
S_L^0 (\Omega)= S_L(\Omega_{SQL})\cdot\frac{1}{2}\left[1+\left(\frac{\Omega}{\Omega_{cav}}\right)^2+\frac{\left(\frac{\Omega_{SQL}}{\Omega}\right)^4}{1+\left(\frac{\Omega}{\Omega_{cav}}\right)^2}\right]  \, .
\end{equation}
Two examples are shown in Fig.~(\ref{cut}), for $\Omega_{cav}= 0.3\,\Omega_{SQL}$ and $\Omega_{cav}= 3\,\Omega_{SQL}$ (dark, thin lines). In any case, the $SQL$ remains the same (as in Eq.~(\ref{SQL})), and at high frequency (above $\Omega_{cav}$ and $\Omega_{SQL}$) the $S_L$ now increases as $\Omega^2$.

For detuned cavities, the most interesting situation is when $\Omega_{cav}$ is around or above $\Omega_{SQL}$. In this case, for $\Omega<\Omega_{SQL}$ the $S_L$ is not significantly modified with respect to the previously studied situation of $\tau \simeq 0$. At high frequency (well above $\Omega_{cav}$) we can write
\begin{equation}
S_L (\Omega)\simeq S_L(\Omega_{SQL})\cdot\frac{1}{2}\frac{1+\Psi^2}{\sin^2(w-\arctan\Psi)}\left(\frac{\Omega}{\Omega_{cav}}\right)^2  \, .
\end{equation}
$S_L$ increases as $\Omega^2$ like in the 'normal case', and is always above it, with an optimal detection phase of $w = \frac{\pi}{2}+\arctan(\Psi)$. What is more peculiar is the behavior at intermediate frequencies. Here a further increase in sensitivity is found (see the 'bump' in the light (read online) dotted curve of Fig.~(\ref{cut})). The $S_L$ is here lower than in the case of negligible $\tau$, yet remaining above the $SQL$.

\section{DUAL detector}

A DUAL gw detector exploits two oscillation modes of a mechanical system with a readout symmetric with respect to the center of mass. Do to the geometry, the responses to the readout force of the two modes must be summed, while the responses to the tidal force of the gw are subtracted. In the frequency region between the two resonance frequencies, the susceptivities of the two modes are in anti-phase, giving a reduced response to the readout force and an enhanced response to the gw. Different configurations have been proposed and studied: two nested spheres~\cite{Cerdonio}, where the relevant modes are the first quadrupolar mode of the inner and outer sphere; two nested cylinders~\cite{Bonaldi}, again acting on the first quadrupolar mode of the nested bodies; a single hollow cylinder, exploiting its first and second quadrupolar modes; a symmetric set of cylinders, where the first DUAL mode is given by the link between them and the second one by their first oscillation mode~\cite{Marin}. 

For our analysis, we consider the simple case of very low (vanishing) first resonance frequency. The susceptivity (i.e., the difference between $\chi_s$ and $\chi_c$) can be written as 
\begin{equation}
\chi(\Omega) = \left[-\frac{1}{\Omega^2}+\frac{1}{\mu (\Omega_R^2-\Omega^2)}\right]\chi_0\Omega_0^2
\end{equation}
where $\Omega_R$ is the (second) DUAL resonance frequency, $\mu$ is an adimensional modal mass and the frequency is normalized to $\Omega_0$. A typical choice of $\Omega_0$ is $\Omega_0 = v_s/R$, where $R$ is a typical dimension of the detector and $v_s$ is the sound velocity in the detector material. $\chi_0$ depends on the material properties and geometrical configuration of the detector.

The response to the gw is
\begin{equation}
H_{GW}(\Omega) = 1+\frac{\Omega^2}{\mu_{GW} (\Omega_R^2-\Omega^2)}
\end{equation}
where $\mu_{GW}$ is an adimensional gw sensitive modal mass. Both $\mu$ and $\mu_{GW}$ are normalized to the detector physical mass and can be calculated from the detector geometry: $\mu$ is a surface overlap integral between the radiation pressure of the field interrogating the surface and the considered mechanical mode; $\mu_{GW}$ is a volume overlap integral between the quadrupolar gw force and the mechanical mode. We take for example in this article the 'QUAD' detector described in Ref.~\cite{Marin}, with four parallel molybdenum cylinders. In this case, we have $\mu = 0.28$, $\mu_{GW} = 1.1$, $\Omega_R = 1.4$.

\begin{figure}[t]
\begin{center}
\includegraphics*[width=1.0\columnwidth]{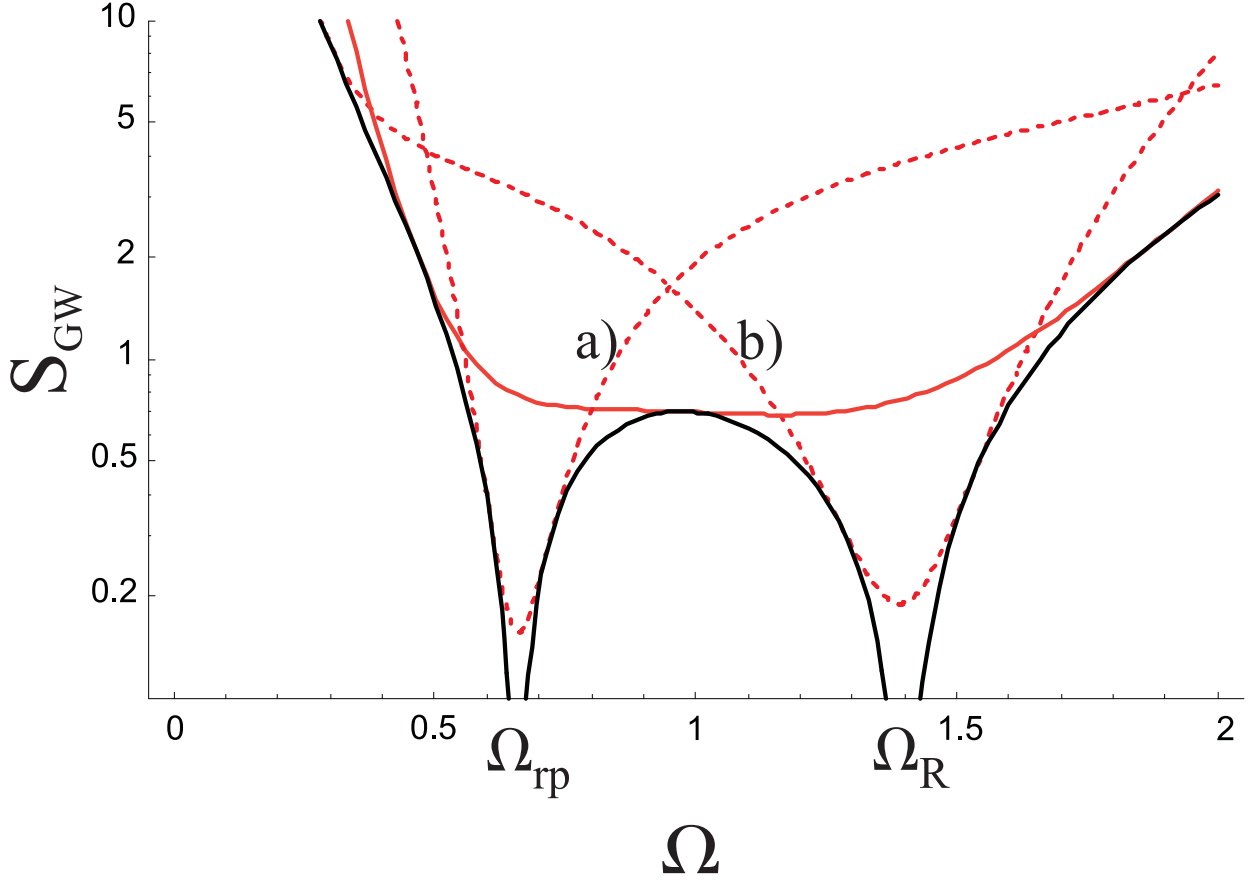}
\end{center}
\caption{Sensitivity to a gravitational wave $S_{GW}$ of a DUAL detector, as a function of the frequency $\Omega$, in arbitrary units. Dark, solid line (black online): $SQL_{GW}$. Light solid line (red online): `normal case` (resonant cavities, $w=\pi/2$) with $p=0.4$. Dotted lines: `normal case` with $p=2$ (a), $p=0.1$ (b).}
\label{dual1}
\end{figure}

Being interested in the sensitivity to a gw signal, we can define it as
\begin{equation}
S_{GW}(\Omega) = \frac{S_L(\Omega)}{|H_{GW}(\Omega)|^2} \, .
\end{equation}
In the case of the free-falling masses MFP considered above, we have $H_{GW}=1$ and therefore $S_{GW}=S_L$. The standard quantum limit is defined similarly to Eq.~(\ref{SQL}):
\begin{equation}
SQL_{GW}= \frac{2 \hbar|\chi(\Omega)|}{\sqrt{\Gamma_m}|H_{GW}(\Omega)|^2}
\end{equation}
and is shown in Fig.~(\ref{dual1}) with a dark, solid line. 

In the 'normal case', varying the input power gives a family of sensitivity curves, limited by the $SQL_{GW}$. A proper choice of the power gives a rather flat curve. For lower power, the sensitivity is peaked around $\Omega_R$ where both the gw signal and the radiation pressure are amplified. A deep in the radiation pressure effect is found around a particular frequency $\Omega_{rp}$ where the interference between the two detector modes gives $\chi(\Omega_{rp})=0$ (for the parameters here considered it happens at $\Omega_{rp}\simeq0.655$). For high input power, around this frequency we have the best sensitivity (Fig.~\ref{dual1}).

\begin{figure}[t]
\begin{center}
\includegraphics*[width=1.0\columnwidth]{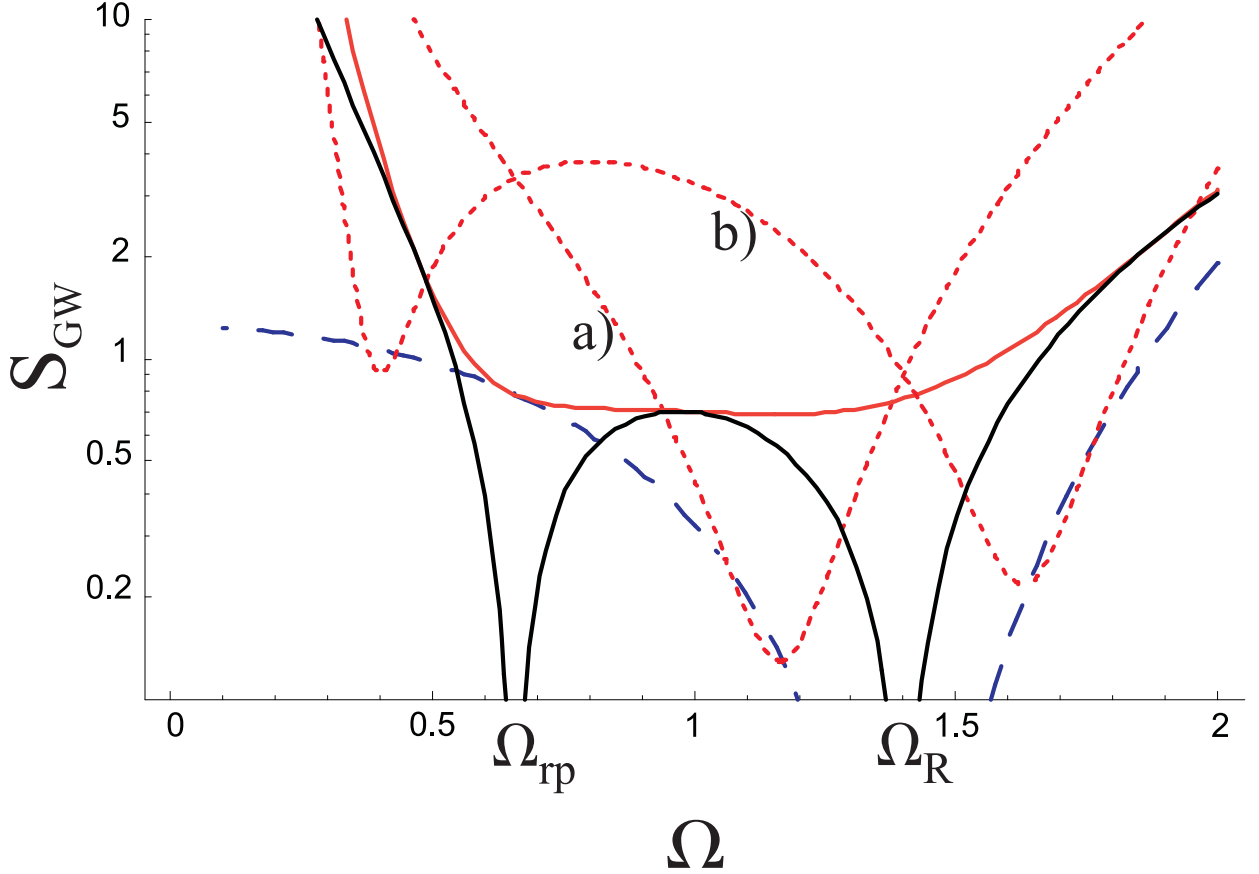}
\end{center}
\caption{Sensitivity to a gravitational wave $S_{GW}$ of a DUAL detector, as a function of the frequency $\Omega$, in arbitrary units. Dark, solid line (black online): $SQL_{GW}$, the other curves are for $p=0.4$; dashed line: resonant case limit; light solid line (red online): `normal case'; dotted lines: $w=\pi/2$ and $\Psi=-0.5$ (a), $\Psi=0.5$ (b).}
\label{dual2}
\end{figure}

Also for a DUAL detector, keeping the cavities at resonance and varying $w$ one  obtains a set of curves whose envelop, shown in Fig.~(\ref{dual2}) with a dashed line, is well below the $SQL_{GW}$ and allows to widen the  detector sensitivity around $\Omega_R$. Changing the detuning $\Psi$, we find a $S_{GW}$ getting even lower. As shown in Fig.~(\ref{dual2}), depending on the parameters, the reduction below $SQL_{GW}$ can occur either in a range between $\Omega_{rp}$ and $\Omega_R$ (where $\chi>0$), or below $\Omega_{rp}$ and above $\Omega_R$ (where $\chi<0$). It can be shown as a general property that, for fixed detuning and phase $w$, it is not possible to obtain a sensitivity curve falling below the $SQL$ both in frequency regions with positive $\chi$ and in regions with negative $\chi$.

\section{Conclusions}

We have presented the analysis of a Michelson-Fabry-Perot interferometer with the addition of the free choice of the detected field quadrature. Our study fully accounts for quantum noise, including radiation pressure effects and possible losses in the cavities, and we consider in particular the case of detuned optical cavities. 

The use of a susceptivity matrix allows to extend the applicability of our results from the usual, free-falling mirrors to the most general mechanical system, including any kind of DUAL detectors with mirrors installed on elastic, mechanically coupled test masses. In view of the development of extended numerical evaluations, considering construction tolerances and variations of system parameters, we have given complete general expressions of the expected output signals.

A physical analysis of simplified expressions shows that, thanks to the possible choice of the detected quadrature, the sensitivity can surpass the SQL even with resonant cavities. This is due to the complete cancellation of the radiation pressure fluctuations in the detected output field quadrature. With detuned cavities, in the case of real susceptivity and loss-less cavities, the peak spectral sensitivity is unlimited if the susceptivity and/or the laser power are strong enough: the ultimate sensitivity is only limited by the reactive component of the susceptivity. We also show that reasonable optical losses do not critically modify the performance.

The scheme that we consider, with the choice of the detected field quadrature, is easier to be implemented than other proposed and tested configurations (e.g., the resonant sideband extraction\cite{Heinzel} or the signal-recycling cavity\cite{Buonanno,Harms}). It must be remarked, however, that here the sensitivity enhancement occurs in a region of high susceptivity (corresponding in the case of free-falling mirrors to the low-frequency region). 

While, for a chosen value of the detection phase, the sensitivity is optimized at a particular frequency (i.e., for a particular value of the susceptivity), one can implement a frequency-dependent rotation of the output field before detection, as proposed and analyzed in several works\cite{Vyatchanin95,Kimble01}, in order to enlarge the useful bandwidth. Without discussing and detailing the possible experimental schemes suitable at this purpose, we give the obtainable sensitivity curves that are indeed broad and deep with respect to the SQL.

The use of detuned cavities is again somehow simpler than other schemes that are based on additional mirrors and cavities or on the use of squeezed light, yet allowing to observe interesting noise reduction effects similar to those reported in the literature concerning the mentioned configurations. We remark that detuned cavities imply in general critical stability problems due, e.g., to radiation pressure and photo-thermal effects\cite{bel,Marino07}. The analysis of this topic is beyond the purpose of this article, and it should be developed for each particular system including the full frequency response and the possible active stabilization.

\section{Appendix A}

The equations for the field and cavity phase linearized fluctuations in the Fourier space can be written in a compact matrix form. Defining an input fluctuations vector $A_{in}(\Omega)$ and a cavity fluctuations vector $A(\Omega)$ as 
\begin{equation}
A(\Omega)= \left( \begin{array}{c}
\delta\tilde{\alpha}_1(\Omega)  \\
\delta\tilde{\alpha}_1^*(\Omega)  \\
\delta\tilde{\psi}_1(\Omega)\\
\delta\tilde{\alpha}_2(\Omega)  \\
\delta\tilde{\alpha}_2^*(\Omega)  \\
\delta\tilde{\psi}_2(\Omega)  \end{array}
\right);\;\;A_{in}(\Omega)=\left( \begin{array}{c}
\sqrt{\gamma_{m1}}(\delta\tilde{\alpha}_{in}-\delta\tilde{\alpha}_V)+\sqrt{2
\gamma_{l1}}\delta\tilde{\alpha}_{l1}  \\
\sqrt{\gamma_{m1}}(\delta\tilde{\alpha}^*_{in}-\delta\tilde{\alpha}^*_V)+\sqrt{2
\gamma_{l1}}\delta\tilde{\alpha}^*_{l1}  \\
\delta\tilde{\psi}_{01}\\
\sqrt{\gamma_{m2}}(\delta\tilde{\alpha}_{in}+\delta\tilde{\alpha}_V)+\sqrt{2
\gamma_{l2}}\delta\tilde{\alpha}_{l2}  \\
\sqrt{\gamma_{m2}}(\delta\tilde{\alpha}^*_{in}+\delta\tilde{\alpha}^*_V)+\sqrt{2
\gamma_{l2}}\delta\tilde{\alpha}^*_{l2}  \\
\delta\tilde{\psi}_{02}
\end{array}\right)
\end{equation}
equations (\ref{dadt}-\ref{ain2}), (\ref{psi}) can be written as
\begin{equation}
\label{eqM}
\mathbf{M}(\Omega)\cdot A(\Omega)=A_{in}(\Omega)
\end{equation}
where
\begin{equation}
\mathbf{M}(\Omega)=
\left(\begin{array}{cccccc} -i\,\tau \,\Omega - i\bar{\psi}_1 + \gamma_1 & 0 & - i\,{{{\bar{\alpha}}}_1}
       & 0 & 0 & 0 \cr 0 & - i\,\tau \,\Omega    + {{{i\bar{\psi} }}_1} + {{\gamma }_1} & i\,
   {{{{{\bar{\alpha}}}_1}}^*} & 0 & 0 & 0 \cr 4\,k^2\,\hbar \,{{\chi }_{11}}\,{{{{{\bar{\alpha}}}_1}}^*} & 4\,k^2\,\hbar \,
   {{\chi }_{11}}\,{{{\bar{\alpha}}}_1} & 1 & 4\,k^2\,\hbar \,{{\chi }_{12}}\,{{{{{\bar{\alpha}}}_2}}^*} & 4\,k^2\,\hbar \,
   \chi_{12}\,{{{\bar{\alpha}}}_2} & 0 \cr 0 & 0 & 0 & - i\,\tau \,\Omega    - {{{i\bar{\psi} }}_2} + 
   {{\gamma }_2} & 0 & - i\,{{{\bar{\alpha}}}_2}   \cr 0 & 0 & 0 & 0 & - i\,\tau \,\Omega    + 
   {{{i\bar{\psi} }}_2} + {{\gamma }_2} & i\,{{{{{\bar{\alpha}}}_2}}^*} \cr 4\,k^2\,\hbar \,{{\chi }_{21}}\,
   {{{\bar{\alpha}}}_1} & 4\,k^2\,\hbar \,{{\chi }_{21}}\,{{{{{\bar{\alpha}}}_1}}^*} & 0 & 4\,k^2\,\hbar \,\chi_{22}\,
   {{{{{\bar{\alpha}}}_2}}^*} & 4\,k^2\,\hbar \,{{\chi }_{22}}\,{{{\bar{\alpha}}}_2} & 1 \cr   
\end{array}\right)\\
\end{equation}

It is useful to write the field fluctuations in terms of amplitude and phase quadratures, defined as
\begin{equation}
\delta p=\delta\tilde{\alpha}+\delta\tilde{\alpha}^* \, \,; \,\, \, \delta q=i(\delta\tilde{\alpha}^*-\delta\tilde{\alpha}) \, .
\end{equation}
Eq.~(\ref{eqM}) becomes
\begin{equation}
\label{sistemaN}
\mathbf{N}(\Omega)\cdot X(\Omega)=X_{in}(\Omega)
\end{equation}
with fluctuation vectors $X(\Omega)$ and $X_{in}(\Omega)$ and system matrix $\mathbf{N}(\Omega)$ given by
\begin{equation}
\label{Xin}
X(\Omega)= \left(
\begin{array}{c}
{{{\delta p}}_1}\\{{{\delta q}}_1}\\
  {{{\delta \tilde{\psi} }}_1}\\{{{\delta p}}_2}\\
  {{{\delta q}}_2}\\{{{\delta \tilde{\psi} }}_2}\end{array}
\right);\;\;X_{in}(\Omega)=\left( \begin{array}{c} 
{\sqrt{{{2\gamma }_{l1}}}} {{{\delta p}}_{{l1}}} +
   {\sqrt{{{\gamma }_{m1}}}} \left( {{{\delta p}}_{{in}}} -
      {{{\delta p}}_V} \right) \\
   {\sqrt{{{2\gamma }_{l1}}}} {{{\delta q}}_{{l1}}} +
   {\sqrt{{{\gamma }_{m1}}}} \left( {{{\delta q}}_{{in}}} -
      {{{\delta q}}_V} \right) \\
  {{{\delta \tilde{\psi} }}_{{01}}}\\
   {\sqrt{{{2\gamma }_{l2}}}} {{{\delta p}}_{{l2}}} +
   {\sqrt{{{\gamma }_{m2}}}} \left( {{{\delta p}}_{{in}}} +
      {{{\delta p}}_V} \right) \\
   {\sqrt{{{2\gamma }_{l2}}}} {{{\delta q}}_{{l2}}} +
   {\sqrt{{{\gamma }_{m2}}}} \left( {{{\delta q}}_{{in}}} +
      {{{\delta q}}_V} \right) \\
  {{{\delta \tilde{\psi} }}_{{02}}} 
\end{array}\right)
\end{equation}
\begin{equation}
\mathbf{N}(\Omega)=\left(\begin{matrix}  -i  \tau_1  \Omega  + {{\gamma }_1} & {{\bar{\psi}
}_1} & \frac{2 {{\bar{\alpha} }_{ {in}}}
     {\sqrt{{{\gamma }_{ {m1}}}}} {{\bar{\psi} }_1}}{{{{\gamma }_1}}^2 + {{{\bar{\psi} }_1}}^2} & 0 & 0 & 0 \cr -{{\bar{\psi} }_1} &
    -i  \tau_1  \Omega  + {{\gamma }_1} & \frac{-2 {{\bar{\alpha} }_{ {in}}} {{\gamma }_1}
     {\sqrt{{{\gamma }_{ {m1}}}}}}{{{{\gamma }_1}}^2 + {{{\bar{\psi} }_1}}^2} & 0 & 0 & 0 \cr \frac{-4 k^2 \hbar
     {{\bar{\alpha} }_{ {in}}} {{\gamma }_1} {\sqrt{{{\gamma }_{ {m1}}}}} {{\chi }_1}}{{{{\gamma }_1}}^2 +
     {{{\bar{\psi} }_1}}^2} & \frac{-4 k^2 \hbar  {{\bar{\alpha} }_{ {in}}} {\sqrt{{{\gamma }_{ {m1}}}}} {{\chi }_1}
     {{\bar{\psi} }_1}}{{{{\gamma }_1}}^2 + {{{\bar{\psi} }_1}}^2} & 1 & \frac{-4 k^2 \hbar  {{\bar{\alpha} }_{ {in}}} {{\gamma }_2}
     {\sqrt{{{\gamma }_{ {m2}}}}} {{\chi }_{12}}}{{{{\gamma }_2}}^2 + {{{\bar{\psi} }_2}}^2} & \frac{-4 k^2 \hbar
     {{\bar{\alpha} }_{ {in}}} {\sqrt{{{\gamma }_{ {m2}}}}} {{\chi }_{12}} {{\bar{\psi} }_2}}{{{{\gamma }_2}}^2 +
     {{{\bar{\psi} }_2}}^2} & 0 \cr 0 & 0 & 0 & -i  \tau_2  \Omega  + {{\gamma }_2} & {{\bar{\psi} }_2} & \frac{2
     {{\bar{\alpha} }_{ {in}}} {\sqrt{{{\gamma }_{ {m2}}}}} {{\bar{\psi} }_2}}{{{{\gamma }_2}}^2 + {{{\bar{\psi} }_2}}^2} \cr 0 &
   0 & 0 & -{{\bar{\psi} }_2} & -i  \tau_2  \Omega  + {{\gamma }_2} & \frac{-2 {{\bar{\alpha} }_{ {in}}} {{\gamma }_2}
     {\sqrt{{{\gamma }_{ {m2}}}}}}{{{{\gamma }_2}}^2 + {{{\bar{\psi} }_2}}^2} \cr \frac{-4 k^2 \hbar
     {{\bar{\alpha} }_{ {in}}} {{\gamma }_1} {\sqrt{{{\gamma }_{ {m1}}}}} {{\chi }_{21}}}{{{{\gamma }_1}}^2 +
     {{{\bar{\psi} }_1}}^2} & \frac{-4 k^2 \hbar  {{\bar{\alpha} }_{ {in}}} {\sqrt{{{\gamma }_{ {m1}}}}}
     {{\chi }_{21}} {{\bar{\psi} }_1}}{{{{\gamma }_1}}^2 + {{{\bar{\psi} }_1}}^2} & 0 & \frac{-4 k^2 \hbar  {{\bar{\alpha} }_{ {in}}}
     {{\gamma }_2} {\sqrt{{{\gamma }_{ {m2}}}}} {{\chi }_2}}{{{{\gamma }_2}}^2 + {{{\bar{\psi} }_2}}^2} & \frac{-4 k^2
     \hbar  {{\bar{\alpha} }_{ {in}}} {\sqrt{{{\gamma }_{ {m2}}}}} {{\chi }_2} {{\bar{\psi} }_2}}{{{{\gamma }_2}}^2 +
     {{{\bar{\psi} }_2}}^2} & 1 \cr
\end{matrix}\right)\\
\end{equation}
From the boundary conditions (\ref{aout}) we obtain the quadrature fluctuations of the reflected fields, according to
\begin{eqnarray}
\delta p_{1}^{out}&=&-\frac{{\delta p_{in}-\delta
p_V}}{\sqrt{2}}+\sqrt{2\gamma_{m1}}\delta
p_1\\\delta q_{1}^{out}&=&-\frac{{\delta
q_{in}-\delta
q_V}}{\sqrt{2}}+\sqrt{2\gamma_{m1}}\delta q_1\\
\delta p_{2}^{out}&=&-\frac{{\delta p_{in}+\delta
p_V}}{\sqrt{2}}+\sqrt{2\gamma_{m2}}\delta
p_2\\\delta q_{2}^{out}&=&-\frac{{\delta
q_{in}+\delta
q_V}}{\sqrt{2}}+\sqrt{2\gamma_{m2}}\delta q_2
\label{boundaryphase}
\end{eqnarray}
Finally, at the output port of the beam splitter, field quadratures are recombined according to
\begin{eqnarray}
\label{outBS}
\delta p_{BS}&=& \frac{1}{\sqrt{2}}(-\delta p_{1}^{out}+\delta
p_{2}^{out}\cos{\theta}+\delta q_{2}^{out}\sin{\theta})\\\nonumber \delta
q_{BS}&=& \frac{1}{\sqrt{2}}(-\delta q_{1}^{out}+\delta
q_{2}^{out}\cos{\theta}+\delta p_{2}^{out}\sin{\theta})
\end{eqnarray}
where $\theta$ is the phase shift introduced by the interferometer
unbalance.

\subsection{Symmetric interferometer}

We will solve the above equations in the case of identical cavities and balanced Michelson interferometer. Therefore we replace $\gamma_{m1} = \gamma_{m2} = \gamma_m$, $\gamma_{l1} = \gamma_{l2} = \gamma_l$, $\tau_1 = \tau_2 = \tau$, and we set $\theta = 0$.

The field quadratures fluctuations at the output port of the beam splitter can be expressed by the product between input fluctuation $X_{in}$ and two vectors of coefficients $P_{BS}[\delta x_i]$ and $Q_{BS}[\delta x_i]$, where $\delta x_i$ is the generic input fluctuation, with $i=$1 to 10: $\delta p_{BS}=P_{BS}\cdot X_{in}$ and $\delta q_{BS}=Q_{BS}\cdot X_{in}$. The vectors $P_{BS}$ and $Q_{BS}$, obtained from Eqs.~(\ref{sistemaN}-\ref{outBS}), are the following:
\begin{eqnarray*}
\Delta \cdot P_{BS}[\delta p_{in}]&=&\frac{1}{{( {\gamma }^2 + {{{\bar{\psi} }_1}}^2 )
}^2{( {\gamma }^2 + {{{\bar{\psi} }_2}}^2 ) }^2}\Big\{ -64 k^4 \gamma
{\hbar }^2 {{{\bar{\alpha} }_{ {in}}}}^4 {{{\gamma }_m}}^3
    ( {{\chi }_1} {{\chi }_2} -
     {{\chi }_{12}} {{\chi }_{21}}  )  {{\bar{\psi} }_1} {{\bar{\psi} }_2}
    ( {{{\bar{\psi} }_1}}^2 - {{{\bar{\psi} }_2}}^2  )  \\\nonumber &+&
        8 k^2 \hbar  {{{\bar{\alpha} }_{ {in}}}}^2 {{{\gamma }_m}}^2
    \Big[
     {{\chi }_1} {{\bar{\psi} }_1} { ( {\gamma }^2 + {{{\bar{\psi} }_2}}^2  ) }^2
       [  ( \gamma  - i  \tau  \Omega   )  {{{\bar{\psi} }_1}}^2 +
        \gamma   (  ( \gamma  - i  \tau  \Omega   )   ( 2 \gamma  - i  \tau  \Omega   )  +
           {{{\bar{\psi} }_2}}^2  )   ]   \\\nonumber &-&{{\chi }_2} {{\bar{\psi} }_2} { ( {\gamma }^2 + {{{\bar{\psi} }_1}}^2  ) }^2
       [   \gamma  {{{\bar{\psi} }_1}}^2    +
         ( \gamma  - i  \tau  \Omega   )
          ( 2 {\gamma }^2 - i  \gamma  \tau  \Omega  + {{{\bar{\psi} }_2}}^2  )   ]   \Big]\\\nonumber &+&
  8 k^2    \hbar  {{{\bar{\alpha} }_{ {in}}}}^2 {{{\gamma }_m}}^2
    ( {\gamma }^2 + {{{\bar{\psi} }_1}}^2  )   ( {\gamma }^2 + {{{\bar{\psi} }_2}}^2  )( 2 \gamma  - i  \tau  \Omega   )
    [ {{\chi }_{12}} {{\bar{\psi} }_1}
       ( \gamma   ( \gamma  - i  \tau  \Omega   )  + {{{\bar{\psi} }_2}}^2  )
          \\\nonumber &-&  {{\chi }_{21}}  {{\bar{\psi} }_2} ( \gamma   ( \gamma  - i  \tau  \Omega   )  + {{{\bar{\psi} }_1}}^2  )
          ] \\\nonumber &+&
    {{\gamma }_m}( \gamma  - i  \tau  \Omega   )  { ( {\gamma }^2 + {{{\bar{\psi} }_1}}^2  ) }^2{ ( {\gamma }^2 + {{{\bar{\psi} }_2}}^2  ) }^2
    ( {{{\bar{\psi} }_1}}^2 - {{{\bar{\psi} }_2}}^2  )  \Big\}
\end{eqnarray*}
\begin{eqnarray*}
\Delta \cdot P_{BS}[\delta p_{V}]&=&\frac{1}{{( {\gamma }^2 + {{{\bar{\psi} }_1}}^2 )
}^2{( {\gamma }^2 + {{{\bar{\psi} }_2}}^2 )
}^2}\times\\\nonumber&\times&\Big\{ -64 k^4 {\hbar}^2 {{{\bar{\alpha}
}_{ {in}}}}^4 {{{\gamma }_m}}^2
    ( {{\chi }_1} {{\chi }_2} - {{\chi }_{12}} {{\chi }_{21}}  )  {{\bar{\psi} }_1} {{\bar{\psi} }_2}
    [  ( {\gamma }^2 + {{{\bar{\psi} }_1}}^2  )   ( {\gamma }^2 + {{{\bar{\psi} }_2}}^2  )  \\\nonumber &+&
     \gamma  {{\gamma }_m}  ( 2 {\gamma }^2 + {{{\bar{\psi} }_1}}^2 + {{{\bar{\psi} }_2}}^2  )   ]\\\nonumber &+&
  8 k^2 \hbar  {{{\bar{\alpha} }_{ {in}}}}^2 {{\gamma }_m}
    \Big[ {{\chi }_2}{{\bar{\psi} }_2} { ( {\gamma }^2 + {{{\bar{\psi} }_1}}^2  ) }^2
       [ - \gamma  {{\gamma }_m}  ( { ( \gamma  - i  \tau  \Omega   ) }^2 + {{{\bar{\psi} }_1}}^2  )
               +  ( -{ ( \gamma  - i  \tau  \Omega   ) }^2 \\\nonumber &+&
            ( \gamma  - i  \tau  \Omega   )  {{\gamma }_m} - {{{\bar{\psi} }_1}}^2  )
          ( {\gamma }^2 + {{{\bar{\psi} }_2}}^2  )   ]  -
     {{\chi }_1} {{\bar{\psi} }_1} { ( {\gamma }^2 + {{{\bar{\psi} }_2}}^2  ) }^2
       [  ( {\gamma }^2 + {{{\bar{\psi} }_1}}^2  )
          ( { ( \gamma  - i  \tau  \Omega   ) }^2 + {{{\bar{\psi} }_2}}^2  ) \\\nonumber &-&
        {{\gamma }_m}  ( i \gamma  \tau  \Omega   (   \gamma  -i \tau  \Omega   )  +
            ( \gamma  - i  \tau  \Omega   )  {{{\bar{\psi} }_1}}^2 - \gamma  {{{\bar{\psi} }_2}}^2  )   ]   \Big] \\\nonumber&+&
     8 k^2 \hbar
 {{{\bar{\alpha} }_{ {in}}}}^2 {{{\gamma }_m}}^2
 ( {\gamma }^2 + {{{\bar{\psi} }_1}}^2  ) ( {\gamma }^2 + {{{\bar{\psi} }_2}}^2  )
   ( 2 \gamma  - i  \tau  \Omega   ) [  {{\chi }_{21}} {{\bar{\psi} }_2}
       ( \gamma   ( \gamma  - i  \tau  \Omega   )  + {{{\bar{\psi} }_1}}^2  )
       \\\nonumber &+&
      {{\chi }_{12}} {{\bar{\psi} }_1}
       ( \gamma   ( \gamma  - i  \tau  \Omega   )  + {{{\bar{\psi} }_2}}^2  )   ]   \\\nonumber &-&
    { ( {\gamma }^2 + {{{\bar{\psi} }_1}}^2  ) }^2
   { ( {\gamma }^2 + {{{\bar{\psi} }_2}}^2  ) }^2
    \Big[  ( { ( \gamma  - i  \tau  \Omega   ) }^2 + {{{\bar{\psi} }_1}}^2  )
       ( { ( \gamma  - i  \tau  \Omega   ) }^2 + {{{\bar{\psi} }_2}}^2  )  \\\nonumber &-&
       {{\gamma }_m} ( \gamma  - i  \tau  \Omega   )
       [2 { ( \gamma  - i  \tau  \Omega   ) }^2 + {{{\bar{\psi} }_1}}^2 + {{{\bar{\psi} }_2}}^2  ]   \Big]  \Big\}
\end{eqnarray*}
\begin{eqnarray*}
\Delta \cdot P_{BS}[\delta q_{in}]&=&\frac{1}{{( {\gamma }^2 + {{{\bar{\psi} }_1}}^2 )
}^2{( {\gamma }^2 + {{{\bar{\psi} }_2}}^2 )
}^2}\times\\\nonumber&\times&\Big\{ -64k^4{\hbar }^2{{{\bar{\alpha}
}_{{in}}}}^4{{{\gamma }_m}}^3
   ( {{\chi }_1}{{\chi }_2} - {{\chi }_{12}}{{\chi }_{21}} ) {{\bar{\psi} }_1} {{\bar{\psi} }_2}
   ( {{\bar{\psi} }_1} - {{\bar{\psi} }_2} )(  {{\bar{\psi} }_1}{{\bar{\psi} }_2} -{\gamma }^2 )
   \\\nonumber&-&
  8k^2\hbar {{{\bar{\alpha} }_{{in}}}}^2{{{\gamma }_m}}^2
   \Big[ {{\chi }_1}{{\bar{\psi} }_1}{( {\gamma }^2 + {{{\bar{\psi} }_2}}^2 ) }^2
      [ {\gamma }^2{{\bar{\psi} }_2} + {{{\bar{\psi} }_1}}^2{{\bar{\psi} }_2} -
        {{\bar{\psi} }_1}( {( \gamma  - i \tau \Omega  ) }^2 + {{{\bar{\psi} }_2}}^2 )  ]\\\nonumber&-&
        {{\chi }_2}{{\bar{\psi} }_2}{( {\gamma }^2 + {{{\bar{\psi} }_1}}^2 ) }^2
        [ {\gamma }^2{{\bar{\psi} }_1} + {{\bar{\psi} }_1}{{{\bar{\psi} }_2}}^2-    {{\bar{\psi} }_2} ( {( \gamma  - i \tau \Omega  ) }^2 + {{{\bar{\psi} }_1}}^2 )
         ]
   \Big] \\\nonumber&-&
  8i k^2\hbar {{{\bar{\alpha} }_{{in}}}}^2
   {{{\gamma }_m}}^2\tau \Omega ( {{\chi }_{12}} - {{\chi }_{21}} ) {{\bar{\psi} }_1}  {{\bar{\psi} }_2}( {\gamma }^2 + {{{\bar{\psi} }_1}}^2 )
 ( {\gamma }^2 + {{{\bar{\psi} }_2}}^2 ) ( 2\gamma  - i \tau \Omega  )  \\\nonumber&+&
  {{\gamma }_m}{( {\gamma }^2 + {{{\bar{\psi} }_1}}^2 ) }^2 {( {\gamma }^2 + {{{\bar{\psi} }_2}}^2 ) }^2( {{\bar{\psi} }_1} - {{\bar{\psi} }_2} )
   ( {( \gamma  - i \tau \Omega  ) }^2 - {{\bar{\psi} }_1}{{\bar{\psi} }_2} )
  \Big\}
\end{eqnarray*}
\begin{eqnarray*}
\Delta \cdot P_{BS}[\delta q_{V}]&=&\frac{1}{{( {\gamma }^2 + {{{\bar{\psi} }_1}}^2 )
}^2{( {\gamma }^2 + {{{\bar{\psi} }_2}}^2 )
}^2}\times\\\nonumber&\times&\Big\{-64 k^4 {\hbar }^2 {{{\bar{\alpha}
}_{ {in}}}}^4 {{{\gamma }_m}}^3
    ( {{\chi }_1} {{\chi }_2} - {{\chi }_{12}} {{\chi }_{21}}  )  {{\bar{\psi} }_1} {{\bar{\psi} }_2}
    ( {{\bar{\psi} }_1} + {{\bar{\psi} }_2}  )   ( {\gamma }^2 + {{\bar{\psi} }_1} {{\bar{\psi} }_2}  ) \\\nonumber &-&
  8 k^2 \hbar  {{{\bar{\alpha} }_{ {in}}}}^2 {{{\gamma }_m}}^2
    \Big[{{\chi }_1} {{\bar{\psi} }_1} { ( {\gamma }^2 + {{{\bar{\psi} }_2}}^2  ) }^2
       [ {\gamma }^2 {{\bar{\psi} }_2} + {{{\bar{\psi} }_1}}^2 {{\bar{\psi} }_2} +
        {{\bar{\psi} }_1}  ( { ( \gamma  - i  \tau  \Omega   ) }^2 + {{{\bar{\psi} }_2}}^2  )   ]\\\nonumber &+&
         {{\chi }_2} {{\bar{\psi} }_2} { ( {\gamma }^2 + {{{\bar{\psi} }_1}}^2  ) }^2
       [ {\gamma }^2 {{\bar{\psi} }_1}+ {{\bar{\psi} }_1} {{{\bar{\psi} }_2}}^2  +
         {{\bar{\psi} }_2} ( { ( \gamma  - i  \tau  \Omega   ) }^2 + {{{\bar{\psi} }_1}}^2  )
        ]
        \Big] \\\nonumber &-&
  8 i  k^2     \hbar  {{{\bar{\alpha} }_{ {in}}}}^2
   {{{\gamma }_m}}^2  \tau  \Omega ( {{\chi }_{12}} + {{\chi }_{21}}  )  {{\bar{\psi} }_1}    {{\bar{\psi} }_2}
   ( {\gamma }^2 +{{{\bar{\psi} }_1}}^2  )
 ( {\gamma }^2 + {{{\bar{\psi} }_2}}^2  ) ( 2 \gamma  - i  \tau  \Omega   ) \\\nonumber &-&
  {{\gamma }_m}   { ( {\gamma }^2 + {{{\bar{\psi} }_2}}^2  ) }^2 { ( {\gamma }^2 + {{{\bar{\psi} }_1}}^2  ) }^2  ( {{\bar{\psi} }_1} + {{\bar{\psi} }_2}  )
    ( { ( \gamma  - i  \tau  \Omega   ) }^2 + {{\bar{\psi} }_1} {{\bar{\psi} }_2}  )
 \Big\}
\end{eqnarray*}
\begin{eqnarray*}
\Delta \cdot P_{BS}[\delta p_{l1}]&=&\frac{ {\sqrt{\gamma  - {{\gamma
}_m}}} {\sqrt{2\gamma_m}} ( {\gamma }^2 + {{{\bar{\psi} }_2}}^2 )} {{(
{\gamma }^2 + {{{\bar{\psi} }_1}}^2 ) }^2{( {\gamma }^2 + {{{\bar{\psi} }_2}}^2
) }^2}\Big\{64 k^4 \gamma  {\hbar }^2 {{{\bar{\alpha} }_{ {in}}}}^4
{{{\gamma }_m}}^2
    ( {{\chi }_1} {{\chi }_2} - {{\chi }_{12}} {{\chi }_{21}}  )  {{\bar{\psi} }_1} {{\bar{\psi} }_2}  \\\nonumber &+&
  8 k^2 \hbar  {{{\bar{\alpha} }_{ {in}}}}^2 {{\gamma }_m}
    [  {{\chi }_1}  {{\bar{\psi} }_1} \gamma  ( {\gamma }^2 + {{{\bar{\psi} }_2}}^2  )
       ( { ( \gamma  - i  \tau  \Omega   ) }^2 + {{{\bar{\psi} }_2}}^2  )
       -
       {{\chi }_2} {{\bar{\psi} }_2}    ( {\gamma }^2 + {{{\bar{\psi} }_1}}^2  )^2 ( \gamma  - i  \tau  \Omega   )
       ] \\\nonumber &-&
   8 k^2   \hbar  {{{\bar{\alpha} }_{ {in}}}}^2 {{\gamma }_m}
   {{\chi }_{21}} {{\bar{\psi} }_2}  ( {\gamma }^2 + {{{\bar{\psi} }_1}}^2  )( 2 \gamma  - i  \tau  \Omega   )
    ( \gamma   ( \gamma  - i  \tau  \Omega   )  + {{{\bar{\psi} }_1}}^2  ) \\\nonumber &-&
   ( {\gamma }^2 + {{{\bar{\psi} }_2}}^2  ) { ( {\gamma }^2 + {{{\bar{\psi} }_1}}^2  ) }^2  ( { ( \gamma  - i  \tau  \Omega   ) }^3
   +
      ( \gamma  - i  \tau  \Omega   )  {{{\bar{\psi} }_2}}^2  ) \Big\}
\end{eqnarray*}
\begin{equation*}
P_{BS}[\delta p_{l2}]\,=\,-P_{BS}[\delta p_{l1}] \,\,\,\,\,\,\,\,\, (1\,\leftrightarrow \,2)    
\end{equation*}
\begin{eqnarray*}
\Delta \cdot P_{BS}[\delta q_{l1}]&=&\frac{ {\sqrt{\gamma  - {{\gamma
}_m}}} {\sqrt{2\gamma_m}} {{\bar{\psi} }_1}( {\gamma }^2 + {{{\bar{\psi}
}_2}}^2 )}
  {{( {\gamma }^2 + {{{\bar{\psi} }_1}}^2 ) }^2{(
{\gamma }^2 + {{{\bar{\psi} }_2}}^2 ) }^2}\Big\{64 k^4 {\hbar }^2
{{{\bar{\alpha} }_{ {in}}}}^4 {{{\gamma }_m}}^2
    ( {{\chi }_1} {{\chi }_2} - {{\chi }_{12}} {{\chi }_{21}}  )  {{\bar{\psi} }_1} {{\bar{\psi} }_2}\\\nonumber &+&
  8 k^2 \hbar  {{{\bar{\alpha} }_{ {in}}}}^2 {{\gamma }_m}
    [ {{\chi }_1} {{\bar{\psi} }_1} ( {\gamma }^2 + {{{\bar{\psi} }_2}}^2  )
       ( { ( \gamma  - i  \tau  \Omega   ) }^2 + {{{\bar{\psi} }_2}}^2  )   +
     {{\chi }_2} {{\bar{\psi} }_2} { ( {\gamma }^2 + {{{\bar{\psi} }_1}}^2  ) }^2  ] \\\nonumber &+&
  8 i  k^2 \hbar  {{{\bar{\alpha} }_{ {in}}}}^2 \tau  \Omega
   {{\gamma }_m} {{\chi }_{21}} {{\bar{\psi} }_2} ( {\gamma }^2 + {{{\bar{\psi} }_1}}^2  )
   ( 2 \gamma  - i  \tau  \Omega   )  \\\nonumber &+&
   ( {\gamma }^2 + {{{\bar{\psi} }_2}}^2  ) { ( {\gamma }^2 + {{{\bar{\psi} }_1}}^2  ) }^2
    ( { ( \gamma  - i  \tau  \Omega   ) }^2 + {{{\bar{\psi} }_2}}^2
      )  \Big\}
\end{eqnarray*}
\begin{equation*}
P_{BS}[\delta q_{l2}]\,=\,-P_{BS}[\delta q_{l1}] \,\,\,\,\,\,\,\,\, (1\,\leftrightarrow \,2)     
\end{equation*}
\begin{eqnarray*}
\Delta \cdot P_{BS}[\delta \psi_{01}]&=&\frac{2 \bar{\alpha}_{in}\gamma_m}
  {{( {\gamma }^2 + {{{\bar{\psi} }_1}}^2 ) }{(
{\gamma }^2 + {{{\bar{\psi} }_2}}^2 ) }}\Big\{8 k^2    \hbar  {{{\bar{\alpha}
}_{ {in}}}}^2 {{\gamma }_m}
    ( {{\chi }_2} + {{\chi }_{21}}  )  {{\bar{\psi} }_1} {{\bar{\psi} }_2}( 2 \gamma - i
 \tau  \Omega   ) \\\nonumber &+&
    {{\bar{\psi} }_1} ( 2 \gamma  - i  \tau  \Omega   )  ( {\gamma }^2 + {{{\bar{\psi} }_2}}^2  )
    ( { ( \gamma  - i  \tau  \Omega   ) }^2 + {{{\bar{\psi} }_2}}^2  )\Big\}
\end{eqnarray*}
\begin{equation*}
P_{BS}[\delta \psi_{02}]\,=\,-P_{BS}[\delta \psi_{01}] \,\,\,\,\,\,\,\,\, (1\,\leftrightarrow \,2)   
\end{equation*}
\begin{eqnarray*}
\Delta \cdot Q_{BS}[\delta p_{in}]&=&\frac{1}{{( {\gamma }^2 + {{{\bar{\psi} }_1}}^2 )
}^2{( {\gamma }^2 + {{{\bar{\psi} }_2}}^2 ) }^2}\Big\{
  64 k^4 {\gamma }^2 {\hbar }^2 {{{\bar{\alpha} }_{ {in}}}}^4 {{{\gamma }_m}}^3
    ( {{\chi }_1} {{\chi }_2} -
    {{\chi }_{12}} {{\chi }_{21}}  )
    \\\nonumber &\times&[ {{\bar{\psi} }_1}({\gamma }^2  + {{{\bar{\psi} }_1}}^2) - {{\bar{\psi} }_2}  ( {\gamma }^2 + {{{\bar{\psi} }_2}}^2  )   ]  \\\nonumber &+&
    8 k^2 \hbar  {{{\bar{\alpha} }_{ {in}}}}^2 {{{\gamma }_m}}^2
    \Big[ {{\chi }_2}  [ {\gamma }^2  ( {\gamma }^2 + {{{\bar{\psi} }_1}}^2
    )^2            ( { ( \gamma  - i  \tau  \Omega   ) }^2 + {{{\bar{\psi} }_1}}^2  )  \\\nonumber &-&
          {{\bar{\psi} }_1}  {{\bar{\psi} }_2}{ ( {\gamma }^2 + {{{\bar{\psi} }_1}}^2  ) }^2  ( {\gamma }^2 + {{{\bar{\psi} }_2}}^2  )
           ]     - {{\chi }_1}  [    {\gamma }^2  ( {\gamma }^2 + {{{\bar{\psi} }_2}}^2
           )^2
          ( { ( \gamma  - i  \tau  \Omega   ) }^2 + {{{\bar{\psi} }_2}}^2  )    \\\nonumber &-&{{\bar{\psi} }_1} {{\bar{\psi} }_2} ( {\gamma }^2 + {{{\bar{\psi} }_1}}^2  )
         { ( {\gamma }^2 + {{{\bar{\psi} }_2}}^2  ) }^2]
       \Big]\\\nonumber &+&
          8 k^2 \hbar  {{{\bar{\alpha} }_{ {in}}}}^2 {{{\gamma }_m}}^2( {\gamma }^2 + {{{\bar{\psi} }_1}}^2  )
           ( {\gamma }^2 + {{{\bar{\psi} }_2}}^2  )
  [ {{\chi }_{21}}
       ( \gamma   ( \gamma  - i  \tau  \Omega   )  + {{{\bar{\psi} }_1}}^2  )
       ( \gamma   ( \gamma  - i  \tau  \Omega   )  - {{{\bar{\psi} }_2}}^2  )
        \\\nonumber &-&
     {{\chi }_{12}}
      ( \gamma   ( \gamma  - i  \tau  \Omega   )  - {{{\bar{\psi} }_1}}^2  )
       ( \gamma   ( \gamma  - i  \tau  \Omega   )  + {{{\bar{\psi} }_2}}^2  )   ]  \\\nonumber &+&
{{\gamma }_m} { ( {\gamma }^2 + {{{\bar{\psi} }_1}}^2  ) }^2  { ( {\gamma
}^2 + {{{\bar{\psi} }_2}}^2  ) }^2 ( {{\bar{\psi} }_1} - {{\bar{\psi} }_2} )
    ( -{ ( \gamma  - i  \tau  \Omega   ) }^2 + {{\bar{\psi} }_1} {{\bar{\psi} }_2}  )
   \Big\}
\end{eqnarray*}
\begin{eqnarray*}
\Delta \cdot Q_{BS}[\delta p_{V}]&=&\frac{1}{{( {\gamma }^2 + {{{\bar{\psi} }_1}}^2 )
}^2{( {\gamma }^2 + {{{\bar{\psi} }_2}}^2 ) }^2}\Big\{
  64 k^4 {\gamma }^2 {\hbar }^2 {{{\bar{\alpha} }_{ {in}}}}^4 {{{\gamma }_m}}^3
    ( {{\chi }_1} {{\chi }_2} - {{\chi }_{12}} {{\chi }_{21}}  )
    \\\nonumber &\times&[ {{\bar{\psi} }_1}  ( {\gamma }^2 + {{{\bar{\psi} }_1}}^2  )  +
     {{\bar{\psi} }_2}  ( {\gamma }^2 + {{{\bar{\psi} }_2}}^2  )   ] \\\nonumber &+&
     8 k^2 \hbar  {{{\bar{\alpha} }_{ {in}}}}^2 {{{\gamma }_m}}^2
    \Big[ {{\chi }_2}  [ {\gamma }^2 { ( {\gamma }^2 + {{{\bar{\psi} }_1}}^2  ) }^2
          ( { ( \gamma  - i  \tau  \Omega   ) }^2 + {{{\bar{\psi} }_1}}^2  )  +
        {{\bar{\psi} }_1} {{\bar{\psi} }_2}{ ( {\gamma }^2 + {{{\bar{\psi} }_1}}^2  ) }^2   ( {\gamma }^2 + {{{\bar{\psi} }_2}}^2  )
         ]  \\\nonumber &+& {{\chi }_1}  [
        {\gamma }^2 { ( {\gamma }^2 + {{{\bar{\psi} }_2}}^2  ) }^2
          ( { ( \gamma  - i  \tau  \Omega   ) }^2 + {{{\bar{\psi} }_2}}^2  )+{{\bar{\psi} }_1} {{\bar{\psi} }_2} ( {\gamma }^2 + {{{\bar{\psi} }_1}}^2  )
         { ( {\gamma }^2 + {{{\bar{\psi} }_2}}^2  ) }^2    ]   \Big]\\\nonumber &-&
  8 k^2 \hbar  {{{\bar{\alpha} }_{ {in}}}}^2 {{{\gamma }_m}}^2
    [   {{\chi }_{21}}  ( {\gamma }^2 + {{{\bar{\psi} }_1}}^2  ) ( {\gamma }^2 + {{{\bar{\psi} }_2}}^2  )
         ( \gamma   ( \gamma  - i  \tau  \Omega   )  + {{{\bar{\psi} }_1}}^2  )
         ( \gamma   ( \gamma  - i  \tau  \Omega   )  - {{{\bar{\psi} }_2}}^2  )
            \\\nonumber &+&
     {{\chi }_{12}}  ( {\gamma }^2 + {{{\bar{\psi} }_1}}^2  )   ( {\gamma }^2 + {{{\bar{\psi} }_2}}^2  )
     ( \gamma   ( \gamma  - i  \tau  \Omega   )  - {{{\bar{\psi} }_1}}^2  )
       ( \gamma   ( \gamma  - i  \tau  \Omega   )  + {{{\bar{\psi} }_2}}^2  )   ]  \\\nonumber &+&
  {{\gamma }_m} { (
{\gamma }^2 + {{{\bar{\psi} }_1}}^2  ) }^2  { ( {\gamma }^2 + {{{\bar{\psi}
}_2}}^2  ) }^2 ( {{\bar{\psi} }_1} + {{\bar{\psi} }_2} )
    ( { ( \gamma  - i  \tau  \Omega   ) }^2 + {{\bar{\psi} }_1} {{\bar{\psi} }_2}  )
   \Big\}
\end{eqnarray*}
\begin{eqnarray*}
\Delta \cdot Q_{BS}[\delta q_{in}]&=&\frac{1}{{( {\gamma }^2 + {{{\bar{\psi} }_1}}^2 )
}^2{( {\gamma }^2 + {{{\bar{\psi} }_2}}^2 ) }^2}\Big\{64 k^4 \gamma
{\hbar }^2 {{{\bar{\alpha} }_{ {in}}}}^4 {{{\gamma }_m}}^3
    ( {{\chi }_1} {{\chi }_2} - {{\chi }_{12}} {{\chi }_{21}}  )  {{\bar{\psi} }_1} {{\bar{\psi} }_2}
    ( {{{\bar{\psi} }_1}}^2 - {{{\bar{\psi} }_2}}^2  )  \\\nonumber &+&
    8 k^2 \hbar  {{{\bar{\alpha} }_{ {in}}}}^2 {{{\gamma }_m}}^2
    \Big[ {{\chi }_1} {{\bar{\psi} }_1} { ( {\gamma }^2 + {{{\bar{\psi} }_2}}^2  ) }^2
       [
         ( \gamma  - i  \tau  \Omega   )
         (i  \gamma  \tau  \Omega+ {{{\bar{\psi} }_1}}^2) - \gamma  {{{\bar{\psi} }_2}}^2  ] \\\nonumber &+&
     {{\chi }_2}{{\bar{\psi} }_2} { ( {\gamma }^2 + {{{\bar{\psi} }_1}}^2  ) }^2
     [ \gamma  {{{\bar{\psi} }_1}}^2 -  ( \gamma  - i  \tau  \Omega   )
          ( i  \gamma  \tau  \Omega  + {{{\bar{\psi} }_2}}^2  )   ]   \Big] \\\nonumber &+&
  8 i  k^2 \tau  \Omega  \hbar  {{{\bar{\alpha} }_{ {in}}}}^2 {{{\gamma }_m}}^2
    ( {\gamma }^2 + {{{\bar{\psi} }_1}}^2  )   ( {\gamma }^2 + {{{\bar{\psi} }_2}}^2  )
    [ {{\chi }_{12}} {{\bar{\psi} }_2} ( \gamma   ( \gamma  - i  \tau  \Omega   )  - {{{\bar{\psi} }_1}}^2  )\\\nonumber&-&
    {{\chi }_{21}} {{\bar{\psi} }_1}  ( \gamma   ( \gamma  - i  \tau  \Omega   )
      -        {{{\bar{\psi} }_2}}^2  )   ]  +
   {{\gamma }_m} { ( {\gamma }^2 + {{{\bar{\psi} }_1}}^2  ) }^2
   { ( {\gamma }^2 + {{{\bar{\psi} }_2}}^2  ) }^2  ( \gamma  - i  \tau  \Omega   )
    ( {{{\bar{\psi} }_1}}^2 - {{{\bar{\psi} }_2}}^2  )  \Big\}
\end{eqnarray*}
\begin{eqnarray*}
\Delta \cdot Q_{BS}[\delta q_{V}]&=&\frac{1}{{( {\gamma }^2 + {{{\bar{\psi} }_1}}^2 )
}^2{( {\gamma }^2 + {{{\bar{\psi} }_2}}^2 )
}^2}\times\\\nonumber&\times&\Big\{
         -64 k^4 {\hbar }^2 {{{\bar{\alpha} }_{ {in}}}}^4 {{{\gamma }_m}}^2
    ( {{\chi }_1} {{\chi }_2} - {{\chi }_{12}} {{\chi }_{21}}  )  {{\bar{\psi} }_1} {{\bar{\psi} }_2}
    \\\nonumber &\times&[  ( {\gamma }^2 + {{{\bar{\psi} }_1}}^2  )   ( {\gamma }^2 + {{{\bar{\psi} }_2}}^2  )
    -
     \gamma  {{\gamma }_m}  ( 2 {\gamma }^2 + {{{\bar{\psi} }_1}}^2 + {{{\bar{\psi} }_2}}^2  )   ]\\\nonumber &+&
  8 k^2 \hbar  {{{\bar{\alpha} }_{ {in}}}}^2 {{\gamma }_m}
    \Big[ -  {{\chi }_2}{{\bar{\psi} }_2} { ( {\gamma }^2 + {{{\bar{\psi} }_1}}^2  ) }^2
         [  ( {\gamma }^2 + {{{\bar{\psi} }_2}}^2  ) ( { ( \gamma  - i  \tau  \Omega   ) }^2 + {{{\bar{\psi} }_1}}^2  )
             \\\nonumber &-&
          {{\gamma }_m}  ( \gamma  {{{\bar{\psi} }_1}}^2 +
              ( \gamma  - i  \tau  \Omega   )
               ( 2 {\gamma }^2 - i  \gamma  \tau  \Omega  + {{{\bar{\psi} }_2}}^2  )   )   ]    \\\nonumber &-&
     {{\chi }_1} {{\bar{\psi} }_1} { ( {\gamma }^2 + {{{\bar{\psi} }_2}}^2  ) }^2
       [  ( {\gamma }^2 + {{{\bar{\psi} }_1}}^2  )
          ( { ( \gamma  - i  \tau  \Omega   ) }^2 + {{{\bar{\psi} }_2}}^2  )  \\\nonumber &-&
        {{\gamma }_m}  ( \gamma  {{{\bar{\psi} }_1}}^2   +
            ( \gamma  - i  \tau  \Omega   )
             ( 2 {\gamma }^2 - i  \gamma  \tau  \Omega  + {{{\bar{\psi} }_2}}^2  )   )  -
           i  \tau  \Omega   ( {{{\bar{\psi} }_1}}^2 - {{{\bar{\psi} }_2}}^2  ) ]   \Big] \\\nonumber &+&
     8 i  k^2 \tau  \Omega
 \hbar  {{{\bar{\alpha} }_{ {in}}}}^2 {{{\gamma }_m}}^2
    ( {\gamma }^2 + {{{\bar{\psi} }_1}}^2  )   ( {\gamma }^2 + {{{\bar{\psi} }_2}}^2  )
    [ {{\chi }_{12}}  {{\bar{\psi} }_2}  ( \gamma   ( \gamma  - i  \tau  \Omega   )  - {{{\bar{\psi} }_1}}^2  )
     \\\nonumber &+& {{\chi }_{21}} {{\bar{\psi} }_1}  ( \gamma   ( \gamma  - i  \tau  \Omega   )  -
        {{{\bar{\psi} }_2}}^2  )   ] \\\nonumber &-&
  { ( {\gamma }^2 + {{{\bar{\psi} }_1}}^2  ) }^2 { ( {\gamma }^2 + {{{\bar{\psi} }_2}}^2  ) }^2
    [  ( { ( \gamma  - i  \tau  \Omega   ) }^2 + {{{\bar{\psi} }_1}}^2  )
       ( { ( \gamma  - i  \tau  \Omega   ) }^2 + {{{\bar{\psi} }_2}}^2  )  \\\nonumber &-&
      {{\gamma }_m}( \gamma  - i  \tau  \Omega   )
       ( 2 { ( \gamma  - i  \tau  \Omega   ) }^2 + {{{\bar{\psi} }_1}}^2 + {{{\bar{\psi} }_2}}^2  )   ]\Big\}
\end{eqnarray*}
\begin{eqnarray*}
\Delta \cdot Q_{BS}[\delta p_{l1}]&=&\frac{ {\sqrt{\gamma  - {{\gamma
}_m}}} {\sqrt{2\gamma_m}} ( {\gamma }^2 + {{{\bar{\psi} }_2}}^2 )} {{(
{\gamma }^2 + {{{\bar{\psi} }_1}}^2 ) }^2{( {\gamma }^2 + {{{\bar{\psi} }_2}}^2
) }^2}\Big\{-64 k^4 {\gamma }^2 {\hbar }^2 {{{\bar{\alpha} }_{ {in}}}}^4
{{{\gamma }_m}}^2
    ( {{\chi }_1} {{\chi }_2} - {{\chi }_{12}} {{\chi }_{21}}  )  {{\bar{\psi} }_2}\\\nonumber &-&
    8 k^2 \hbar  {{{\bar{\alpha} }_{ {in}}}}^2 {{\gamma }_m}
    [ {{\chi }_1} {\gamma }^2  ( {\gamma }^2 + {{{\bar{\psi} }_2}}^2  )
       ( { ( \gamma  - i  \tau  \Omega   ) }^2 + {{{\bar{\psi} }_2}}^2  )
       +{{\chi }_2} {{\bar{\psi} }_1} {{\bar{\psi} }_2} { ( {\gamma }^2 + {{{\bar{\psi}
}_1}}^2  ) }^2]   \\\nonumber &+&
  8 k^2 \hbar  {{{\bar{\alpha} }_{ {in}}}}^2 {{\gamma }_m} {{\chi }_{21}}( {\gamma }^2 + {{{\bar{\psi} }_1}}^2  )
    ( \gamma   ( \gamma  - i  \tau  \Omega   )  - {{{\bar{\psi} }_2}}^2  )
     ( \gamma   ( \gamma  - i  \tau  \Omega   )
    +
     {{{\bar{\psi} }_1}}^2  )   \\\nonumber &+&
   {{\bar{\psi} }_1} ( {\gamma }^2 + {{{\bar{\psi} }_2}}^2  )  { ( {\gamma }^2 + {{{\bar{\psi} }_1}}^2  ) }^2
   ( { ( i  \gamma  + \tau  \Omega   ) }^2 - {{{\bar{\psi} }_2}}^2  )
    \Big\}
\end{eqnarray*}
\begin{equation*}
Q_{BS}[\delta p_{l2}]\,=\,-Q_{BS}[\delta p_{l1}] \,\,\,\,\,\,\,\,\, (1\,\leftrightarrow \,2)   
\end{equation*}
\begin{eqnarray*}
\Delta \cdot Q_{BS}[\delta q_{l1}]&=&\frac{ {\sqrt{\gamma  - {{\gamma
}_m}}} {\sqrt{2\gamma_m}} ( {\gamma }^2 + {{{\bar{\psi}
}_2}}^2 )}
  {{( {\gamma }^2 + {{{\bar{\psi} }_1}}^2 ) }^2{(
{\gamma }^2 + {{{\bar{\psi} }_2}}^2 ) }^2}\Big\{-64 k^4 \gamma
 {\hbar }^2 {{{\bar{\alpha} }_{ {in}}}}^4 {{{\gamma }_m}}^2
    ( {{\chi }_1} {{\chi }_2} - {{\chi }_{12}} {{\chi }_{21}}  )  {{\bar{\psi} }_1} {{\bar{\psi} }_2}  \\\nonumber &-&
   8 k^2 \hbar  {{{\bar{\alpha} }_{ {in}}}}^2 {{\gamma }_m}
    [  {{\chi }_1} {{\bar{\psi} }_1} \gamma  ( {\gamma }^2 + {{{\bar{\psi} }_2}}^2  )
       ( { ( \gamma  - i  \tau  \Omega   ) }^2 + {{{\bar{\psi} }_2}}^2  ) + {{\chi }_2}{{\bar{\psi} }_2}  { ( {\gamma }^2 + {{{\bar{\psi} }_1}}^2  ) }^2 ( \gamma  - i  \tau  \Omega   )
    ] \\\nonumber &-&
  8 i  k^2  \hbar  {{{\bar{\alpha} }_{ {in}}}}^2 \tau  \Omega {{\gamma }_m} {{\chi }_{21}} {{\bar{\psi} }_1}
    ( {\gamma }^2 + {{{\bar{\psi} }_1}}^2  )   ( {\gamma }^2 - i  \gamma  \tau  \Omega  - {{{\bar{\psi} }_2}}^2  ) \\\nonumber &-&
       {  ( {\gamma }^2 + {{{\bar{\psi} }_2}}^2  )( {\gamma }^2 + {{{\bar{\psi} }_1}}^2  ) }^2
    [ { ( \gamma  - i  \tau  \Omega   ) }^3 +
      ( \gamma  - i  \tau  \Omega   )  {{{\bar{\psi} }_2}}^2  ]
 \Big\}
\end{eqnarray*}
\begin{equation*}
Q_{BS}[\delta q_{l2}]\,=\,-Q_{BS}[\delta q_{l1}] \,\,\,\,\,\,\,\,\, (1\,\leftrightarrow \,2)   
\end{equation*}
\begin{eqnarray*}
\Delta \cdot Q_{BS}[\delta \psi_{01}]&=&\frac{2 \bar{\alpha}_{in}\gamma_m}
  {{( {\gamma }^2 + {{{\bar{\psi} }_1}}^2 ) }{(
{\gamma }^2 + {{{\bar{\psi} }_2}}^2 ) }}\times\\\nonumber&\times&\Big\{ -8
k^2 \hbar  {{{\bar{\alpha} }_{ {in}}}}^2 {{\gamma }_m}
    [ {{\chi }_2} {{\bar{\psi} }_2} (   \gamma   ( \gamma  - i  \tau  \Omega   )     -{{{\bar{\psi} }_1}}^2  )
          + {{\chi }_{21}} {{\bar{\psi} }_1}
       (   \gamma   ( \gamma  - i  \tau  \Omega   )     - {{{\bar{\psi} }_2}}^2  )   ]\\\nonumber&-&
       ( {\gamma }^2 + {{{\bar{\psi} }_2}}^2  )   ( { ( \gamma  - i  \tau  \Omega   ) }^2 + {{{\bar{\psi} }_2}}^2
      )(   \gamma   ( \gamma -i  \tau  \Omega   )     - {{{\bar{\psi} }_1}}^2)
      \Big\}
\end{eqnarray*}
\begin{equation*}
Q_{BS}[\delta \psi_{02}]\,=\,-Q_{BS}[\delta \psi_{01}] \,\,\,\,\,\,\,\,\, (1\,\leftrightarrow \,2)     
\end{equation*}
where $\Delta$ is the determinant of $\mathbf{N}$, expressed by 
\begin{eqnarray*}
\Delta&=&\frac{1}{(\gamma^2 +
\bar{\psi}_1^2)(\gamma^2 + \bar{\psi}_2^2 )}\Big\{64k^4\hbar^2\bar{\alpha}_{in}^4\gamma_m^2
(\chi_1\chi_2 - \chi_{12}\chi_{21}) 
{{\bar{\psi} }_1}{{\bar{\psi} }_2}\\\nonumber&+&( {\gamma }^2 + {{{\bar{\psi} }_1}}^2 ) 
[ {( \gamma  - i \tau \Omega  ) }^2 +
{{{\bar{\psi} }_1}}^2 ] ( {\gamma }^2 + {{{\bar{\psi} }_2}}^2 ) 
[ {( \gamma  - i \tau \Omega  ) }^2 +
{{{\bar{\psi} }_2}}^2 ]+\\\nonumber&&
8k^2\hbar {{\bar{\alpha} }_{{in}}}^2 \gamma_m
\big[\chi_2 \bar{\psi}_2
( {\gamma }^2 + {{{\bar{\psi} }_1}}^2 ) 
[ {( \gamma  - i \tau \Omega  ) }^2 +
{{{\bar{\psi} }_1}}^2 ]+\\\nonumber&&
{{\chi }_1}{{\bar{\psi} }_1}
( {\gamma }^2 + {{{\bar{\psi} }_2}}^2 ) 
[ {( \gamma  - i \tau \Omega  ) }^2 +
{{{\bar{\psi} }_2}}^2 ]\big]\Big\}.
\end{eqnarray*}
The homodyne detection after the beam splitter allows to choose the output quadrature $\delta E_{out}=\delta p_{BS} \cos{w} + \delta q_{BS} \sin{w} $, according to the detection phase $w$. It is useful to describe also $\delta E_{out}$ as product $V_{out} \cdot X_{in}$, with a coefficients vector $V_{out}$ defined as
\begin{equation}
\label{Vout}
V_{out} = P_{BS} \cos{w} + Q_{BS} \sin{w} \, .
\end{equation} 

The input fluctuations $\delta \psi_{01}$ and $\delta \psi_{02}$ contain the gw signal. They also include thermal noise and any kind of classical fluctuations of the cavities length, that we are not considering in this article since in our approach they are not distinguishable from the gw signal. Taking an optimal interferometer orientation, the gw signal is proportional to $\delta (L_1-L_2) = (2k)^{-1}(\delta\psi_{01}-\delta\psi_{02})$. Therefore we define a sensitivity $S_L(\Omega)$ as
\begin{equation}
S_L(\Omega)=\frac{\Sigma_{i=1,8} |V_{out}[\delta x_i]|^2 \cdot S_{x_i x_i}}{(2k)^2|V_{out}[\delta\psi_{01}]-V_{out}[\delta\psi_{02}]|^2}
\end{equation}
where $S_{x_i x_i}$ are the spectral densities of the input fluctuations $\delta x_i$, which are assumed uncorrelated. We take all noise spectra double-sided and normalized to shot noise. Therefore, for all the vacuum fluctuations (i.e., for $\delta p_V, \delta q_V, \delta p_{l1}, \delta q_{l1}, \delta p_{l2}, \delta q_{l2})$ the spectra are $S_{x_i x_i}=1$. 

In totally symmetric conditions, we have $\bar{\psi}_1=\bar{\psi}_2=\bar{\psi}$, $\chi_{11}=\chi_{22}=\chi_s$ and $\chi_{12}=\chi_{21}=\chi_c$. To write clearer expressions, we define $\Psi = \bar{\psi}/\gamma$ (detuning normalized to the half cavity linewidth); $P_{in} = \hbar kc\bar{\alpha}_{in}^2$ (input laser power); $\Gamma_m = \gamma_m/\gamma$; $\Omega_{cav}=\gamma/\tau$ (cutoff angular frequency of the cavity). The expression of $S_L(\Omega)$ is
\begin{eqnarray}
\label{NSsimm}
&&S_L=
\left\{\sqrt{\Gamma_m}\,p
    \left[ {( 2\Psi \cos{w} + ({\Psi }^2-1)\sin{w} ) }^2 + 
      \left(\frac{\Omega}{\Omega_{cav}}\right)^2(\sin{w} -\Psi \cos{w})^2
        \right] \right\}^{-1}
\\\nonumber&\times&\hbar 
        \Big\{ \frac{p^2}{2}|(\chi_s-\chi_c)|^2 
         \left[ 1 + \frac{3{\Psi }^2-1}{1 + {\Psi }^2} \cos{2w}+ 
           \frac{\Psi ( {\Psi }^2-3)} {1 + {\Psi }^2} \sin{2w} + 
           \frac{{\Psi }^2}{2{{\Gamma }_m}} \right] 
           \\\nonumber &+&  
           \frac{p}{{\sqrt{{{\Gamma }_m}}}}\Big[ {Re(\chi_s-\chi_c) }\Big( \Psi ( 1 + {\Psi }^2 )
               \left(1+{\Psi }^2-\left(\frac{\Omega}{\Omega_{cav}}\right)^2\right)
               \\\nonumber &+& 2\Psi 
               {{{\Gamma }}_m}\left(  2{\Psi }^2 -2 - 
                 \left(\frac{\Omega}{\Omega_{cav}}\right)^2 \right) \cos{2w}  \\\nonumber &+&      
              {{{\Gamma }}_m} 
               \left( 1 + {\Psi }^4 - {\Psi }^2
                  \left(6+\left(\frac{\Omega}{\Omega_{cav}}\right)^2\right)+ 
                 \left(\frac{\Omega}{\Omega_{cav}}\right)^2 \right) \sin{2w}  \Big)  \\\nonumber &+&  
           {Im(\chi_s-\chi_c) }\Big( 
              2\Psi ( 1 + {\Psi }^2 ) \frac{\Omega}
               {{{\Omega }_{{cav}}}} ({{\Gamma }_m} -1  )  \Big)  \Big] \\\nonumber &+&    
      {( 1 + {\Psi }^2 ) }^2\left[ {\Psi }^4 + 
         2{\Psi }^2\left( 1 - \left(\frac{\Omega}{\Omega_{cav}}\right)^2\right)  + 
         \left(1+\left(\frac{\Omega}{\Omega_{cav}}\right)^2 \right)^2 \right]\Big\}.
\end{eqnarray}
We notice that, in this case, $V_{out}[\delta\psi_{01}]=V_{out}[\delta\psi_{02}]$ and the detection is not sensitive to the common mode of the interferometer.
In the short cavity regime, with $\Omega/\Omega_{cav}\simeq 0$, the Eq.~(\ref{NSsimm}) reduces to
\begin{eqnarray}
\label{NS0tau}
S_{L}^{0}&=&\Big\{{\sqrt{{{\Gamma }_m}}}\,p
    \left[2\Psi \cos{w} + ({\Psi }^2-1)\sin{w}
        \right]^2 \Big\}^{-1}\\\nonumber&\times&\hbar 
        \Big\{ \frac{p^2}{2}|(\chi_s-\chi_c)|^2 
         \left[ 1 + \frac{3{\Psi }^2-1}{1 + {\Psi }^2} \cos{2w}+ 
           \frac{\Psi ( {\Psi }^2-3)}{ 1 + {\Psi }^2} \sin{2w} + 
           \frac{{\Psi }^2}{2{{\Gamma }_m}} \right] \\\nonumber &+&  
           \frac{p}{{\sqrt{{{\Gamma }_m}}}}{Re(\chi_s-\chi_c) }\Big[ \Psi {( 1 + {\Psi }^2 ) }^2
                  +   
              {{{\Gamma }}_m} \sin{2w}
               \big( 1 + {\Psi }^4 - 6{\Psi }^2
                 \big)    \\\nonumber &+&  
           4{{{\Gamma }}_m}\Psi ({\Psi }^2-1) \cos{2w}
                 \Big]+
      {( 1 + {\Psi }^2 ) }^4   
       \Big\}  \,\,\,.
 \end{eqnarray}



\begin{thebibliography}{99}
\bibitem{nota1} The authors of Ref.~\cite{Arcizet}, besides a particular choice of detection phase, plot sensitivity curves maintaining a fixed product $P_{in}/(1+\Psi^2)^2$, thus considering an input power increasing with the detuning.
\end{thebibliography}
\end{document}